\documentclass[openacc]{rsproca_new}
\usepackage[utf8]{inputenc}
\usepackage{amsmath, amsthm, amssymb, amsfonts, dsfont, mathrsfs, dirtytalk, tikz-cd, adjustbox, url, nicefrac, booktabs,array, graphicx,todonotes,longtable, tabu,bm,cite,comment,bm,titlesec,hyperref}
\usepackage[T1]{fontenc}    
\usepackage[english]{babel}

\newcommand{\R}{\mathbb R}

\newcommand{\E}{\mathbb E}
\newcommand{\B}{\mathcal B}

\renewcommand{\H}{\operatorname{H}}

\newcommand{\dkl}{\operatorname{D_{K L}}}
\renewcommand{\ker}{\operatorname{k e r}}
\newcommand{\id}{\operatorname{I d}}

\newtheorem{lemma}{\bf Lemma}[section]
\newtheorem{definition}{\bf Definition}[section]
\newtheorem{remark}{\bf Remark}[section]

\newtheorem{example}{\bf Example}[section]

\begin{document}

\title{Bayesian Mechanics for Stationary Processes}

\author{
Lancelot Da Costa$^{1,2}$, Karl Friston$^{2}$, Conor Heins$^{3,4,5}$ and Grigorios A. Pavliotis$^{1}$}

\address{$^{1}$Department of Mathematics, Imperial College London, London SW7 2AZ, UK \\
$^{2}$Wellcome Centre for Human Neuroimaging, University College London, London WC1N 3AR, UK \\
$^{3}$Department of Collective Behaviour, Max Planck Institute of Animal Behavior, Konstanz D-78457, Germany \\
$^{4}$Centre for the Advanced Study of Collective Behaviour, University of Konstanz, Konstanz D-78457, Germany\\
$^{5}$Department of Biology, University of Konstanz, Konstanz D-78457, Germany}

\subject{Stochastic processes, mathematical biology, Bayesian statistics, stochastic control, mathematical modelling, statistical physics}


\corres{Lancelot Da Costa\\
\email{l.da-costa@imperial.ac.uk}}

\begin{abstract}
This paper develops a Bayesian mechanics for adaptive systems.

Firstly, we model the interface between a system and its environment with a Markov blanket. This affords conditions under which states internal to the blanket encode information about external states.

Second, we introduce dynamics and represent adaptive systems as Markov blankets at steady-state. This allows us to identify a wide class of systems whose internal states appear to infer external states, consistent with variational inference in Bayesian statistics and theoretical neuroscience.

Finally, we partition the blanket into sensory and active states. It follows that active states can be seen as performing active inference and well-known forms of stochastic control (such as PID control), which are prominent formulations of adaptive behaviour in theoretical biology and engineering.
\end{abstract}


\begin{fmtext}
\section{Introduction}
Any object of study must be, implicitly or explicitly, separated from its environment. This implies a boundary that separates it from its surroundings, and which persists for at least as long as the system exists.

\end{fmtext}


\maketitle

In this article, we explore the consequences of a boundary mediating interactions between states internal and external to a system. This provides a useful metaphor to think about biological systems, which comprise spatially bounded, interacting components, nested at several spatial scales \cite{hespMultiscaleViewEmergent2019,kirchhoffMarkovBlanketsLife2018}: for example, the membrane of a cell acts as a boundary through which the cell communicates with its environment, and the same can be said of the sensory receptors and muscles that bound the nervous system.

By examining the dynamics of persistent, bounded systems, we identify a wide class of systems wherein the states internal to a boundary appear to infer those states outside the boundary---a description which we refer to as Bayesian mechanics. Moreover, if we assume that the boundary comprises sensory and active states, we can identify the dynamics of active states with well-known descriptions of adaptive behaviour from theoretical biology and stochastic control. 

In what follows, we link a purely mathematical formulation of interfaces and dynamics with descriptions of belief updating and behaviour found in the biological sciences and engineering. Altogether, this can be seen as a model of adaptive agents, as these interface with their environment through sensory and active states and furthermore behave so as to preserve a target steady-state.

\subsection{Outline of paper}

This paper has three parts, each of which introduces a simple, but fundamental, move.
\begin{enumerate}
    \item The first is to partition the world into internal and external states whose boundary is modelled with a Markov blanket \cite{pearlGraphicalModelsProbabilistic1998,bishopPatternRecognitionMachine2006}. This allows us to identify conditions under which internal states encode information about external states.
    \item The second move is to equip this partition with stochastic dynamics. The key consequence of this is that internal states can be seen as continuously inferring external states, consistent with variational inference in Bayesian statistics and with predictive processing accounts of biological neural networks in theoretical neuroscience.
    \item The third move is to partition the boundary into sensory and active states. It follows that active states can be seen as performing active inference and stochastic control, which are prominent descriptions of adaptive behaviour in biological agents, machine learning and robotics.
\end{enumerate}

\subsection{Related work}

The emergence and sustaining of complex (dissipative) structures have been subjects of long-standing research starting from the work of Prigogine \cite{nicolisSelforganizationNonequilibriumSystems1977,goldbeterDissipativeStructuresBiological2018}, followed notably by Haken’s synergetics \cite{hakenSynergeticsIntroductionNonequilibrium1978}, and in recent years, the statistical physics of adaptation \cite{perunovStatisticalPhysicsAdaptation2016}. A central theme of these works is that complex systems can only emerge and sustain themselves far from equilibrium \cite{jefferyStatisticalMechanicsLife2019,englandStatisticalPhysicsSelfreplication2013,skinnerImprovedBoundsEntropy2021}.

Information processing has long been recognised as a hallmark of cognition in biological systems. In light of this, theoretical physicists have identified basic instances of information processing in systems far from equilibrium using tools from information theory, such as how a drive for metabolic efficiency can lead a system to become predictive \cite{dunnLearningInferenceNonequilibrium2013,stillThermodynamicCostBenefit2020,stillThermodynamicsPrediction2012,ueltzhofferThermodynamicsPredictionDissipative2020}.

A fundamental aspect of biological systems is a self-organisation of various interacting components at several spatial scales \cite{hespMultiscaleViewEmergent2019,kirchhoffMarkovBlanketsLife2018}. Much research currently focuses on multipartite processes---modelling interactions between various sub-components that form biological systems---and how their interactions constrain the thermodynamics of the whole \cite{kardesThermodynamicUncertaintyRelations2021,wolpertMinimalEntropyProduction2020,wolpertUncertaintyRelationsFluctuation2020,crooksMarginalConditionalSecond2019,horowitzThermodynamicsContinuousInformation2014}.

At the confluence of these efforts, researchers have sought to explain cognition in biological systems. Since the advent of the 20th century, Bayesian inference has been used to describe various cognitive processes in the brain \cite{pougetInferenceComputationPopulation2003,knillBayesianBrainRole2004,fristonFreeenergyPrincipleUnified2010,raoPredictiveCodingVisual1999,fristonActionBehaviorFreeenergy2010}. In particular, the free energy principle \cite{fristonFreeenergyPrincipleUnified2010}, a prominent theory of self-organisation from the neurosciences, postulates that Bayesian inference can be used to describe the dynamics of multipartite, persistent systems modelled as Markov blankets at non-equilibrium steady-state \cite{fristonParcelsParticlesMarkov2020,fristonLifeWeKnow2013,parrMarkovBlanketsInformation2020,fristonFreeEnergyPrinciple2019a,fristonStochasticChaosMarkov2021}.

This paper connects and develops some of the key themes from this literature. Starting from fundamental considerations about adaptive systems, we develop a physics of things that hold beliefs about other things--consistently with Bayesian inference--and explore how it relates to known descriptions of action and behaviour from the neurosciences and engineering. Our contribution is theoretical: from a biophysicist's perspective, this paper describes how Bayesian descriptions of biological cognition and behaviour can emerge from standard accounts of physics. From an engineer's perspective this paper contextualises some of the most common stochastic control methods and reminds us how these can be extended to suit more sophisticated control problems.

\subsection{Notation}

Let $\Pi \in \R^{d \times d}$ be a square matrix with real coefficients. Let $\eta, b, \mu$ denote a partition of the states $[\![1, d]\!]$, so that
\begin{align*}
    \Pi = \begin{bmatrix} \Pi_{\eta} & \Pi_{\eta b} & \Pi_{\eta \mu}\\
 \Pi_{b \eta} &\Pi_{b}&\Pi_{b \mu} \\
 \Pi_{\mu \eta}& \Pi_{\mu b}& \Pi_{\mu} \end{bmatrix}.
\end{align*}
We denote principal submatrices with one index only (i.e., we use $\Pi_{\eta}$ instead of $\Pi_{\eta\eta}$). Similarly, principal submatrices involving various indices are denoted with a colon

\begin{align*}
    \Pi_{\eta :b } &:=\begin{bmatrix} \Pi_{\eta} & \Pi_{\eta b} \\
 \Pi_{b \eta} &\Pi_{b} \end{bmatrix}.
\end{align*}

When a square matrix $\Pi$ is symmetric positive-definite we write $\Pi \succ 0$. $\ker$, $\operatorname{Im}$ and $\cdot^-$ respectively denote the kernel, image and Moore-Penrose pseudo-inverse of a linear map or matrix, e.g., a non-necessarily square matrix such as $\Pi_{\mu b}$. In our notation, indexing takes precedence over (pseudo) inversion, for example,
\begin{align*}
    \Pi_{\mu b}^{-} := \left(\Pi_{\mu b}\right)^- \neq (\Pi^-)_{\mu b}.
\end{align*}

\section{Markov blankets}
\label{sec: markov blanket}

\begin{figure}[!h]
\centering\includegraphics[width=0.6\textwidth]{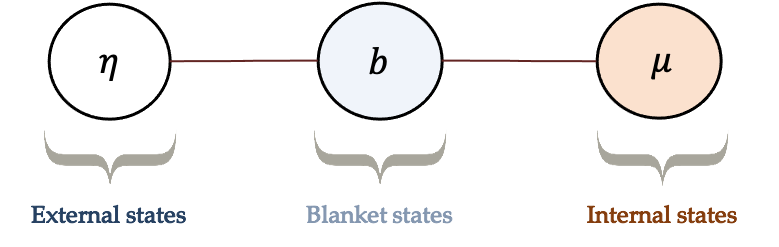}
\caption{\textbf{Markov blanket} depicted graphically as an undirected graphical model, also known as a Markov random field \cite{wainwrightGraphicalModelsExponential2007,bishopPatternRecognitionMachine2006}. (A Markov random field is a Bayesian network whose directed arrows are replaced by undirected arrows). The circles represent random variables. The lines represent conditional dependencies between random variables. The Markov blanket condition means that there is no line between $\mu$ and $\eta$. This means that, $\mu$ and $\eta$ are conditionally independent given $b$. In other words, knowing the internal state $\mu$, does not afford additional information about the external state $\eta$ when the blanket state $b$ is known. Thus blanket states act as an informational boundary between internal and external states.}
\label{fig: 3 way MB}
\end{figure}

The section formalises the notion of boundary between a system and its environment as a Markov blanket \cite{pearlGraphicalModelsProbabilistic1998,bishopPatternRecognitionMachine2006}, depicted graphically in Figure \ref{fig: 3 way MB}. Intuitive examples of a Markov blanket are that of a cell membrane, mediating all interactions between the inside and the outside of the cell, or that of sensory receptors and muscles that bound the nervous system.

To formalise this intuition, we model the world's state as a random variable $x$ with corresponding probability distribution $p$ over a state-space $\mathcal X= \R^d$. We partition the state-space of $x$ into \textit{external}, \textit{blanket} and \textit{internal} states:
\begin{align*}
x&= (\eta, b, \mu)\\
\mathcal X&= \mathcal E \times \mathcal B \times \mathcal I.
\end{align*}
External, blanket and internal state-spaces ($\mathcal E, \mathcal B, \mathcal I$) are taken to be Euclidean spaces for simplicity.

A Markov blanket is a statement of conditional independence between internal and external states given blanket states.

\begin{definition}[Markov blanket]
A Markov blanket is defined as

\begin{equation}
\label{eq: def Markov blanket}\tag{M.B.}
          \eta \perp \mu \mid b
\end{equation}
That is, blanket states are a Markov blanket separating $\mu, \eta$ \cite{pearlGraphicalModelsProbabilistic1998,bishopPatternRecognitionMachine2006}.
\end{definition}

The existence of a Markov blanket can be expressed in several equivalent ways
\begin{align}
\label{eq: Markov blanket densities factorising}
  \eqref{eq: def Markov blanket}  \iff p(\eta, \mu | b) = p(\eta | b)p(\mu | b)
  \iff  p(\eta | b, \mu) = p(\eta | b)   \iff p(\mu | b, \eta) = p(\mu | b).
\end{align}

For now, we will consider a (non-degenerate) Gaussian distribution $p$ encoding the distribution of states of the world
\begin{align*}
    p(x)&:= \mathcal N(x; 0,\Pi^{-1}), \quad \Pi \succ 0,
\end{align*}
with associated precision (i.e., inverse covariance) matrix $\Pi$. Throughout, we will denote the (positive definite) covariance by $\Sigma := \Pi^{-1}$. Unpacking \eqref{eq: Markov blanket densities factorising} in terms of Gaussian densities, we find that a Markov blanket is equivalent to a sparsity in the precision matrix

\begin{align}
\label{eq: MB is sparsity in precision matrix}
    \eqref{eq: def Markov blanket} \iff \Pi_{\eta \mu}= \Pi_{\mu \eta}=0.
\end{align}

\begin{example}
For example,
\begin{align*}
 &\Pi = \begin{bmatrix} 2 & 1 &0\\
 1&2&1 \\
 0& 1& 2\end{bmatrix} \Rightarrow \Sigma_{\eta:b}^{-1} = \begin{bmatrix} 2 & 1 \\
 1&1.5\end{bmatrix}, \Sigma_{b: \mu}^{-1} = \begin{bmatrix} 1.5 & 1 \\
 1& 2\end{bmatrix}
\end{align*}
Then,
\begin{align*}
    p(\eta, \mu |b) &\propto p(\eta, \mu , b)\propto \exp\left( -\frac 1 2 x\cdot \Pi x\right)\\
    &\propto \exp\left( -\frac 1 2 \begin{bmatrix} \eta , b \end{bmatrix}\Sigma_{\eta:b}^{-1} \begin{bmatrix} \eta \\ b \end{bmatrix} - \frac 1 2 \begin{bmatrix} b , \mu\end{bmatrix}\Sigma_{b: \mu}^{-1} \begin{bmatrix}  b\\\mu \end{bmatrix}\right)  \propto p(\eta, b)p( b, \mu ) \propto p(\eta|b)p(\mu |b).
\end{align*}
Thus, the Markov blanket condition \eqref{eq: Markov blanket densities factorising} holds.
\end{example}

\subsection{Expected internal and external states}

Blanket states act as an information boundary between external and internal states. Given a blanket state, we can express the conditional probability densities over external and internal states (using \eqref{eq: Markov blanket densities factorising} and \cite[Prop. 3.13]{eatonMultivariateStatisticsVector2007})\footnote{Note that $\Pi_{\eta},\Pi_{\mu}$ are invertible as principal submatrices of a positive definite matrix.}

\begin{equation}
\label{eq: posterior beliefs}
\begin{split}
    p(\eta|b)&= \mathcal N(\eta; \Sigma_{\eta b} \Sigma_{b}^{-1} b,\: \Pi_{\eta}^{-1}), \\ 
    p(\mu|b)&= \mathcal N(\mu; \Sigma_{\mu b} \Sigma_{b}^{-1} b,\: \Pi_{\mu}^{-1}). 
\end{split}
\end{equation}

This enables us to associate to any blanket state its corresponding expected external and expected internal states:
    
    \begin{align*}
    \boldsymbol \eta(b) &= \E [\eta \mid b]= \E_{p(\eta|b)}[\eta] = \Sigma_{\eta b} \Sigma_{b}^{-1} b \in \mathcal E\\
    \boldsymbol \mu(b) &= \E [\mu \mid b] =\E_{p(\mu|b)}[\mu] =\Sigma_{\mu b} \Sigma_{b}^{-1} b \in \mathcal I.
\end{align*}

Pursuing the example of the nervous system, each sensory impression on the retina and oculomotor orientation (blanket state) is associated with an expected scene that caused sensory input (expected external state) and an expected pattern of neural activity in the visual cortex (expected internal state) \cite{parrComputationalNeurologyActive2019}.

\subsection{Synchronisation map}

A central question is whether and how expected internal states encode information about expected external states. For this, we need to characterise a synchronisation function $\sigma$, mapping the expected internal state to the expected external state, given a blanket state $\sigma(\boldsymbol \mu(b))=\boldsymbol \eta(b)$. This is summarised in the following commutative diagram:

\begin{equation*}
  \begin{tikzcd}
     & b \in \mathcal B \arrow[dl, "\boldsymbol \eta"']\arrow[dr, "\boldsymbol \mu"]  &  \\
    \operatorname{Image}(\boldsymbol \eta) \arrow[rr, leftarrow, dotted, "\sigma"'] &  & \operatorname{Image}(\boldsymbol \mu)
    \end{tikzcd}
\end{equation*}

The existence of $\sigma$ is guaranteed, for instance, if the expected internal state completely determines the blanket state---that is, when no information is lost in the mapping $b \mapsto \boldsymbol \mu(b)$ in virtue of it being one-to-one. In general, however, many blanket states may correspond to an unique expected internal state. Intuitively, consider the various neural pathways that compress the signal arriving from retinal photoreceptors \cite{meisterNeuralCodeRetina1999}, thus many different (hopefully similar) retinal impressions lead to the same signal arriving in the visual cortex.

\subsubsection{Existence}

The key for the existence of a function $\sigma$ mapping expected internal states to expected external states given blanket states, is that for any two blanket states associated with the same expected internal state, these be associated with the same expected external state. This non-degeneracy means that the internal states (e.g., patterns of activity in the visual cortex) have enough capacity to represent all possible expected external states (e.g., 3D scenes of the environment). We formalise this in the following Lemma:

\begin{lemma}
\label{lemma: equiv sigma well defined}
The following are equivalent:
\begin{enumerate}
    \item There exists a function $\sigma : \operatorname{Image}( \boldsymbol \mu)   \to \operatorname{Image} (\boldsymbol \eta)$ such that for any blanket state $b \in \mathcal B$
    \begin{equation*}
        \sigma(\boldsymbol \mu(b))=\boldsymbol \eta(b).
    \end{equation*}
    \item For any two blanket states $b_1, b_2 \in \B$
        \begin{equation*}
        \boldsymbol \mu(b_1)= \boldsymbol \mu(b_2) \Rightarrow \boldsymbol \eta(b_1)= \boldsymbol \eta(b_2).
        \end{equation*}
    \item $\ker \Sigma_{\mu b} \subset \ker \Sigma_{\eta b}$.
    \item $\ker \Pi_{\mu b} \subset \ker \Pi_{\eta b}$.
\end{enumerate}
\end{lemma}

See Appendix \ref{app: sync map existence proof} for a proof of Lemma \ref{lemma: equiv sigma well defined}.

\begin{example}
\begin{itemize}
    \item When external, blanket and internal states are one-dimensional, the existence of a synchronisation map is equivalent to $\Pi_{\mu b} \neq 0$ or $\Pi_{\mu b}= \Pi_{\eta b} =0$.
    \item If $\Pi_{\mu b}$ is chosen at random--its entries sampled from a non-degenerate Gaussian or uniform distribution--then $\Pi_{\mu b}$ has full rank with probability $1$. If furthermore the blanket state-space $\mathcal B$ has lower or equal dimensionality than the internal state-space $\mathcal I$, we obtain that $\Pi_{\mu b}$ is one-to-one (i.e., $\ker \Pi_{\mu b}= 0$) with probability $1$. Thus, in this case, the conditions of Lemma \ref{lemma: equiv sigma well defined} are fulfilled with probability $1$.
\end{itemize}
\end{example}

\subsubsection{Construction}

The key idea to map an expected internal state $\boldsymbol \mu(b)$ to an expected external state $\boldsymbol \eta(b)$ is to: 1) find a blanket state that maps to this expected internal state (i.e., by inverting $\boldsymbol \mu$) and 2) from this blanket state, find the corresponding expected external state (i.e., by applying $\boldsymbol \eta$):

\begin{equation*}
  \begin{tikzcd}
    & & b \in \mathcal B \arrow[ddll, "\boldsymbol \eta"']\arrow[ddrr, "\boldsymbol \mu"]  &  &\\
    &&&& \\
    \operatorname{Image}(\boldsymbol \eta) \arrow[rrrr, leftarrow, "\sigma = \boldsymbol \eta \circ \boldsymbol \mu^-"'] &  & && \operatorname{Image}(\boldsymbol \mu) \arrow[uull, bend left, rightarrow, "\boldsymbol \mu^-" near end]
    \end{tikzcd}
\end{equation*}

We now proceed to solving this problem. Given an internal state $\mu$, we study the set of blanket states $b$ such that $\boldsymbol \mu(b)= \mu$
\begin{equation}
\label{eq: inverse problem blanket states}
    \boldsymbol \mu(b)= \Sigma_{\mu b} \Sigma_{b}^{-1} b= \mu \iff b \in \boldsymbol \mu^{-1} (\mu)= \Sigma_{b}\Sigma_{\mu b}^{-1}\mu.
\end{equation}
Here the inverse on the right hand side of \eqref{eq: inverse problem blanket states} is understood as the preimage of a linear map. We know that this system of linear equations has a vector space of solutions given by \cite{jamesGeneralisedInverse1978} 

\begin{equation}
\label{eq: solution system of equations}
    \boldsymbol \mu^{-1} (\mu)=\left\{\Sigma_{b}\Sigma_{\mu b}^-\mu + \left(\id - \Sigma_{b}\Sigma_{\mu b}^-\Sigma_{\mu b} \Sigma_{b}^{-1}\right)b : b \in \mathcal B\right\}.
\end{equation}
Among these, we choose
\begin{equation*}
    \boldsymbol \mu^- (\mu)= \Sigma_{b}\Sigma_{\mu b}^-\mu.
\end{equation*}

\begin{definition}[Synchronisation map]
We define a synchronisation function that maps to an internal state a corresponding most likely internal state\footnote{This mapping was derived independently of our work in \cite[Section 3.2]{aguileraHowParticularPhysics2021}.}\footnote{Replacing $\boldsymbol \mu^- (\mu)$ by any other element of \eqref{eq: solution system of equations} would lead to the same synchronisation map provided that the conditions of Lemma \ref{lemma: equiv sigma well defined} are satisfied.}
\begin{equation*}
\label{eq: sync map def}
\begin{split}
    \sigma &: \operatorname{Im} \boldsymbol \mu   \to \operatorname{Im} \boldsymbol \eta \\
    \mu &\mapsto \boldsymbol{\eta}(\boldsymbol \mu^- (\mu)) =  \Sigma_{\eta b}\Sigma_{\mu b}^-\mu = \Pi^{-1}_\eta \Pi_{\eta b} \Pi_{\mu b}^- \Pi_\mu \mu.
\end{split}
\end{equation*}
The expression in terms of the precision matrix is a byproduct of Appendix \ref{app: sync map existence proof}.
\end{definition}

\begin{figure}
    \centering
    \includegraphics[width= 0.47\textwidth]{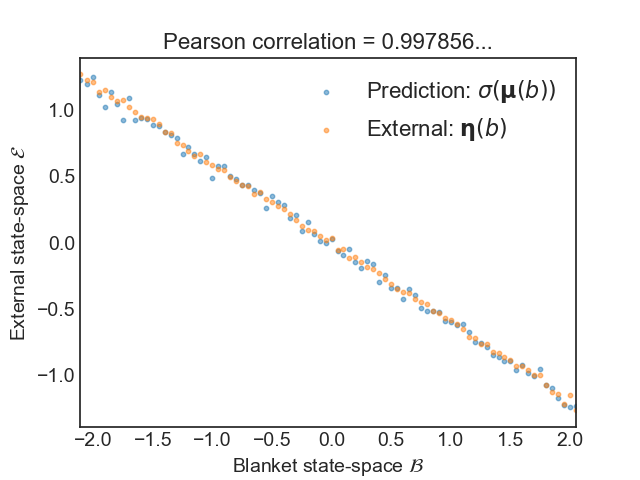}
    \includegraphics[width= 0.47\textwidth]{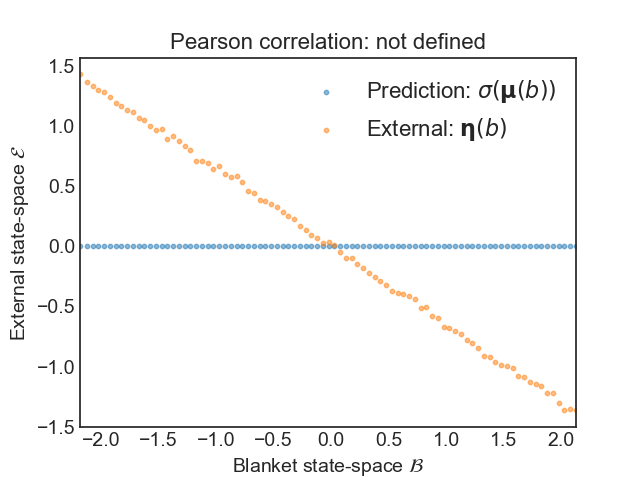}
    \caption{\textbf{Synchronisation map: example and non-example.} This figure plots expected external states given blanket states $\boldsymbol \eta (b)$ (in orange), and the corresponding prediction encoded by internal states $\sigma (\boldsymbol \mu (b))$ (in blue). In this example, external, blanket and internal state-spaces are taken to be one dimensional. We show the correspondence under the conditions of Lemma \ref{lemma: equiv sigma well defined} (left panel) and when these are not satisfied (right panel). To generate these data, 1) we drew $10^6$ samples from a Gaussian distribution with a Markov blanket, 2) we partitioned the blanket state-space into several bins, 3) we obtained the expected external and internal states given blanket states empirically by averaging samples from each bin, and finally, 4) we applied the synchronisation map to the (empirical) expected internal states given blanket states.}
    \label{fig: sigma example non-example}
\end{figure}

Note that we can always define such $\sigma$, however, it is only when the conditions of Lemma \ref{lemma: equiv sigma well defined} are fulfilled that $\sigma$ maps expected internal states to expected external states $\sigma (\boldsymbol \mu (b)) = \boldsymbol \eta (b)$. When this is not the case, the internal states do not fully represent external states, which leads to a partly degenerate type of representation, see Figure \ref{fig: sigma example non-example} for a numerical illustration obtained by sampling from a Gaussian distribution, in the non-degenerate (left) and degenerate cases (right), respectively.

\section{Bayesian mechanics}

In order to study the time-evolution of systems with a Markov blanket, we introduce dynamics into the external, blanket and internal states. Henceforth, we assume a synchronisation map under the conditions of Lemma \ref{lemma: equiv sigma well defined}.

\subsection{Processes at a Gaussian steady-state}

\begin{figure}
    \centering
    \includegraphics[width= 0.7\textwidth]{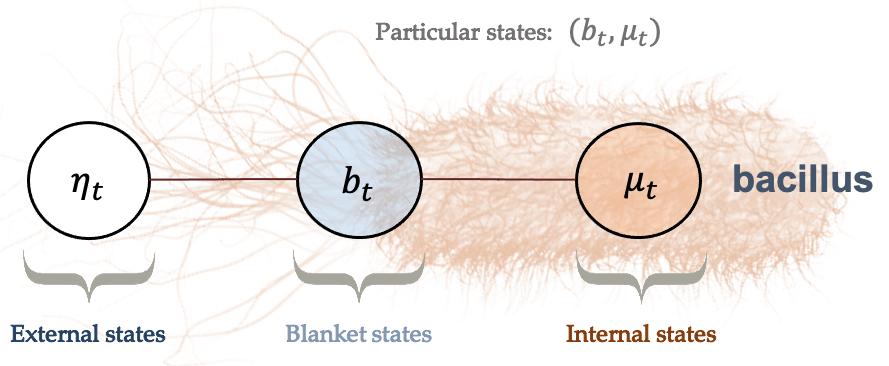}
    \caption{\textbf{Markov blanket evolving in time.} We use a bacillus to depict an intuitive example of a Markov blanket that persists over time. Here, the blanket states represent the membrane and actin filaments of the cytoskeleton, which mediate all interactions between internal states and the external medium (external states).}
    \label{fig: 3 way dynamical MB}
\end{figure}

We consider stochastic processes at a Gaussian steady-state $p$ with a Markov blanket. The steady-state assumption means that the system's overall configuration persists over time (e.g., it does not dissipate). In other words, we have a Gaussian density $p = \mathcal N(0, \Pi^{-1})$ with a Markov blanket \eqref{eq: MB is sparsity in precision matrix} and a stochastic process distributed according to $p$ at every point in time
\begin{equation*}
    x_t \sim p = \mathcal N(0, \Pi^{-1}) \quad \text{ for any } t.
\end{equation*}
Recalling our partition into external, blanket and internal states, this affords a Markov blanket that persists over time, see Figure \ref{fig: 3 way dynamical MB}
\begin{equation}
\label{eq: MB over time}
   x_t= (\eta_t,b_t, \mu_t) \sim p \Rightarrow \eta_t \perp \mu_t \mid b_t.
\end{equation}

\begin{figure}
    \centering
    \includegraphics[width= 0.47\textwidth]{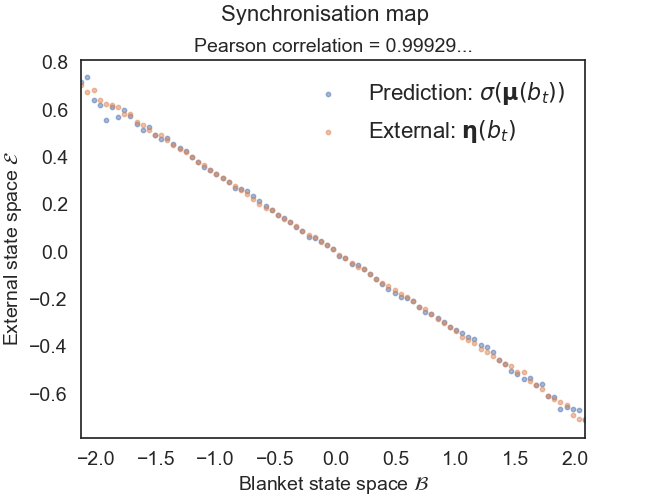}
    \includegraphics[width= 0.47\textwidth]{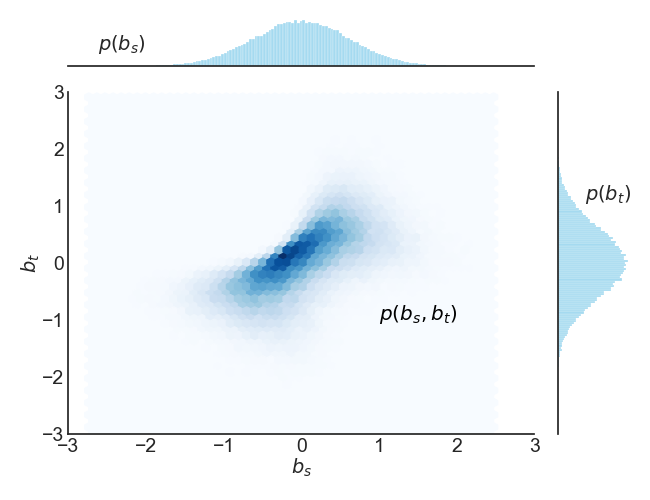}
    \caption{\textbf{Processes at a Gaussian steady-state}. This figure illustrates the synchronisation map and transition probabilities of processes at a Gaussian steady-state. \textit{Left:} We plot the synchronisation map as in Figure \ref{fig: sigma example non-example}, only, here, the samples are drawn from trajectories of a diffusion process \eqref{eq: Gaussian preserving diffusions} with a Markov blanket. Although this is not the case here, one might obtain a slightly noisier correspondence between predictions $\sigma(\boldsymbol \mu(b_t))$ and expected external states $\boldsymbol \eta(b_t)$---compared to Figure \ref{fig: sigma example non-example}---in numerical discretisations of a diffusion process. This is because the steady-state of a numerical discretisation usually differs slightly from the steady-state of the continuous-time process \cite{mattinglyConvergenceNumericalTimeAveraging2010}.
    \textit{Right:} This panel plots the transition probabilities of the same diffusion process  \eqref{eq: Gaussian preserving diffusions}, for the blanket state at two different times. The joint distribution (depicted as a heat map) is not Gaussian but its marginals---the steady-state density---are Gaussian. This shows that in general, processes at a Gaussian steady-state are not Gaussian processes. In fact, the Ornstein-Uhlenbeck process is the only stationary diffusion process \eqref{eq: Gaussian preserving diffusions} that is a Gaussian process, so the transition probabilities of non-linear diffusion processes \eqref{eq: Gaussian preserving diffusions} are never multivariate Gaussians.}
    \label{fig: stationary process with a Markov blanket}
\end{figure}

Note that we do not require $x_t$ to be independent samples from the steady-state distribution $p$. On the contrary, $x_t$ may be generated by extremely complex, non-linear, and possibly stochastic equations of motion. See Example \ref{eg: processes at a gaussian steady-state} and Figure \ref{fig: stationary process with a Markov blanket} for details.

\begin{example}
\label{eg: processes at a gaussian steady-state}
The dynamics of $x_t$ are described by a stochastic process at a Gaussian steady-state $p = \mathcal N(0, \Pi^{-1})$. There is a large class of such processes, for example:

\begin{itemize}
\item Stationary diffusion processes, with initial condition $x_0 \sim p$. Their time-evolution is given by an Itô stochastic differential equation (see Appendix \ref{app: helmholtz decomposition}):
\begin{equation}
\label{eq: Gaussian preserving diffusions}
\begin{split}
    dx_t &= (\Gamma+Q)(x_t) \nabla \log p( x_t) dt + \nabla \cdot (\Gamma+Q)(x_t) dt+ \varsigma(x_t) dW_t, \quad  \\
    &=-(\Gamma+Q)(x_t) \Pi x_t dt +\nabla \cdot (\Gamma+Q)(x_t) dt+ \varsigma (x_t) dW_t\\
    \Gamma&:= \varsigma \varsigma^\top/2 ,\quad  Q= -Q^\top.
\end{split}
\end{equation}
Here, $W_t$ is a standard Brownian motion (a.k.a., Wiener process) \cite{rogersDiffusionsMarkovProcesses2000a, pavliotisStochasticProcessesApplications2014} and $\varsigma,\Gamma,Q$ are sufficiently well-behaved matrix fields (see Appendix \ref{app: helmholtz decomposition}). Namely, $\Gamma$ is the diffusion tensor (half the covariance of random fluctuations), which drives dissipative flow; $Q$ is an arbitrary antisymmetric matrix field which drives conservative (i.e., solenoidal) flow. We emphasise that there are no non-degeneracy conditions on the matrix field $\varsigma$---in particular, the process is allowed to be non-ergodic or even completely deterministic (i.e., $\varsigma \equiv 0$). Also, $\nabla \cdot $ denotes the divergence of a matrix field defined as $\left(\nabla \cdot (\Gamma + Q) \right)_i := \sum_{j} \frac{\partial}{\partial x_j} (\Gamma + Q)_{ij}$.
\item More generally, $x_t$ could be generated by any Markov process at steady-state $p$, such as the zig-zag process or the bouncy particle sampler \cite{bierkensZigZagProcessSuperefficient2019, bierkensPiecewiseDeterministicScaling2017,bouchard-coteBouncyParticleSampler2018}, by any mean-zero Gaussian process at steady-state $p$ \cite{rasmussenGaussianProcessesMachine2004} or by any random dynamical system at steady-state $p$ \cite{arnoldRandomDynamicalSystems1998}.
\end{itemize}
\end{example}

\begin{remark}
When the dynamics are given by an Itô stochastic differential equation \eqref{eq: Gaussian preserving diffusions}, a Markov blanket of the steady-state density \eqref{eq: MB is sparsity in precision matrix} does not preclude reciprocal influences between internal and external states \cite{biehlTechnicalCritiqueParts2021,fristonInterestingObservationsFree2021}. For example,
\begin{align*}
 &\Pi = \begin{bmatrix} 2 & 1 &0\\
 1&2&1 \\
 0& 1& 2\end{bmatrix}, \quad Q \equiv \begin{bmatrix} 0 & 0 &1\\
 0&0&0 \\
 -1& 0& 0\end{bmatrix},\quad  \varsigma \equiv \operatorname{Id}_3 \\
  &\Rightarrow d\begin{bmatrix} \eta_t \\b_t \\ \mu_t \end{bmatrix}=- \begin{bmatrix} 1 & 1.5 &2 \\
  0.5 & 1 &0.5 \\
  -2 &-0.5&1\end{bmatrix} \begin{bmatrix} \eta_t \\b_t \\ \mu_t \end{bmatrix} dt + \varsigma dW_t.
\end{align*}
Conversely, the absence of reciprocal coupling between two states in the drift in some instances, though not always, leads to conditional independence \cite{biehlTechnicalCritiqueParts2021,aguileraHowParticularPhysics2021,fristonStochasticChaosMarkov2021}.
\end{remark}

\subsection{Maximum a posteriori estimation}

The Markov blanket \eqref{eq: MB over time} allows us to harness the construction of Section \ref{sec: markov blanket} to determine expected external and internal states given blanket states
    \begin{align*}
         \boldsymbol \eta_t := \boldsymbol \eta(b_t) \qquad \boldsymbol \mu_t:= \boldsymbol \mu(b_t).
    \end{align*}
Note that $\boldsymbol \eta, \boldsymbol \mu$ are linear functions of blanket states; since $b_t$ generally exhibits rough sample paths, $\boldsymbol \eta_t,\boldsymbol \mu_t$ will also exhibit very rough sample paths.

We can view the steady-state density $p$ as specifying the relationship between external states ($\eta$, causes) and particular states ($b, \mu$, consequences). In statistics, this corresponds to a generative model, a probabilistic specification of how (external) causes generate (particular) consequences.

By construction, the expected internal states encode expected external states via the synchronisation map
\begin{align*}
    \sigma( \boldsymbol \mu_t)&=\boldsymbol \eta_t,
\end{align*}
which manifests a form of generalised synchrony across the Markov blanket \cite{jafriGeneralizedSynchronyCoupled2016,cuminGeneralisingKuramotoModel2007,palaciosEmergenceSynchronyNetworks2019}. Moreover, the expected internal state $\boldsymbol \mu_t$ effectively follows the most likely cause of its sensations
    \begin{align*}
       \sigma( \boldsymbol \mu_t)&= \arg \max p(\eta_t \mid b_t) \quad \text{for any } t.
    \end{align*}
This has an interesting statistical interpretation as expected internal states perform maximum a posteriori (MAP) inference over external states.

\subsection{Predictive processing}

We can go further and associate to each internal state $\mu$ a probability distribution over external states, such that each internal state encodes beliefs about external states
    \begin{align}
        q_\mu(\eta)&:= \mathcal N(\eta; \sigma(\mu), \Pi_{\eta}^{-1}). \label{eq: def approx posterior}
    \end{align}
We will call $q_\mu$ the approximate posterior belief associated with the internal state $\mu$ due to the forecoming connection to inference. Under this specification, the mean of the approximate posterior depends upon the internal state, while its covariance equals that of the true posterior w.r.t. external states \eqref{eq: posterior beliefs}. It follows that the approximate posterior equals the true posterior when the internal state $\mu$ equals the expected internal state $\boldsymbol{\mu}(b)$ (given blanket states):

\begin{equation}
    q_{\mu}(\eta)=p(\eta|b) \iff \mu = \boldsymbol{\mu}(b). \label{eq: approx posterior equals true posterior}
\end{equation}
  
Note a potential connection with epistemic accounts of quantum mechanics; namely, a world governed by classical mechanics ($\sigma \equiv 0$ in \eqref{eq: Gaussian preserving diffusions}) in which each agent encodes Gaussian beliefs about external states could appear to the agents as reproducing many features of quantum mechanics \cite{bartlettReconstructionGaussianQuantum2012}.

Under this specification \eqref{eq: approx posterior equals true posterior}, expected internal states are the unique minimiser of a Kullback-Leibler divergence \cite{kullbackInformationSufficiency1951} 
\begin{align*}
   \boldsymbol \mu_t = \arg \min_\mu \dkl[q_{\mu}(\eta)\| p(\eta|b)]
\end{align*}
that measures the discrepancy between beliefs about the external world $q_\mu(\eta)$ and the posterior distribution over external variables. Computing the KL divergence (see Appendix \ref{app: free energy}), we obtain

\begin{align}
\label{eq: precision weighted prediction error}
  \boldsymbol \mu_t = \arg \min_\mu (\sigma(\mu)-\boldsymbol \eta_t)\Pi_\eta(\sigma(\mu)-\boldsymbol \eta_t)
\end{align}

In the neurosciences, the right hand side of \eqref{eq: precision weighted prediction error} is commonly known as a (squared) precision-weighted prediction error: the discrepancy between the prediction and the (expected) state of the environment is weighted with a precision matrix \cite{bogaczTutorialFreeenergyFramework2017,raoPredictiveCodingVisual1999,fristonPredictiveCodingFreeenergy2009} that derives from the steady-state density. This equation is formally similar to that found in predictive coding formulations of biological function \cite{chaoLargeScaleCorticalNetworks2018,iglesiasHierarchicalPredictionErrors2013,dawModelBasedInfluencesHumans2011,raoPredictiveCodingVisual1999}, which stipulate that organisms minimise prediction errors, and in doing so optimise their beliefs to match the distribution of external states.

\subsection{Variational Bayesian inference}

\begin{figure}
    \centering
    \includegraphics[width= 0.47\textwidth]{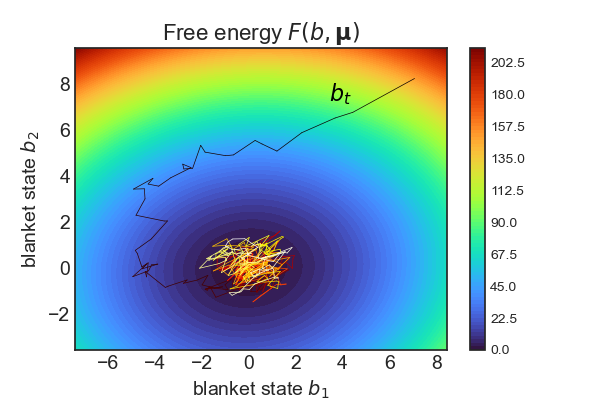}
    \includegraphics[width= 0.47\textwidth]{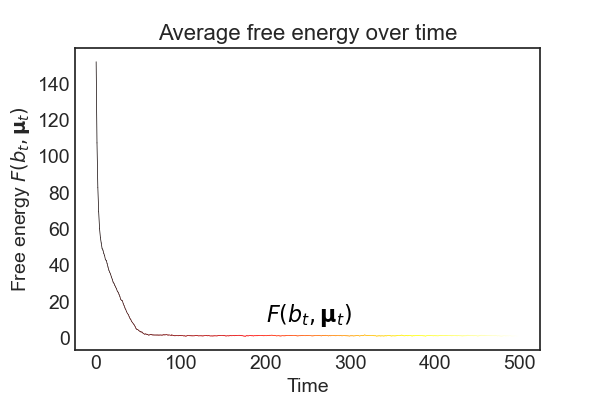}
    \includegraphics[width= 0.47\textwidth]{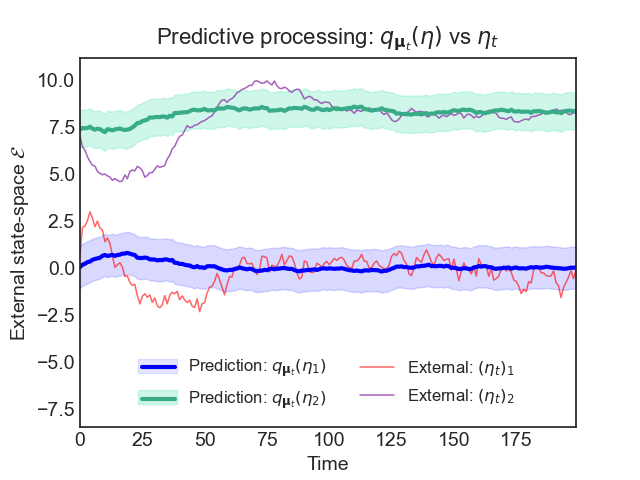}
    \includegraphics[width= 0.47\textwidth]{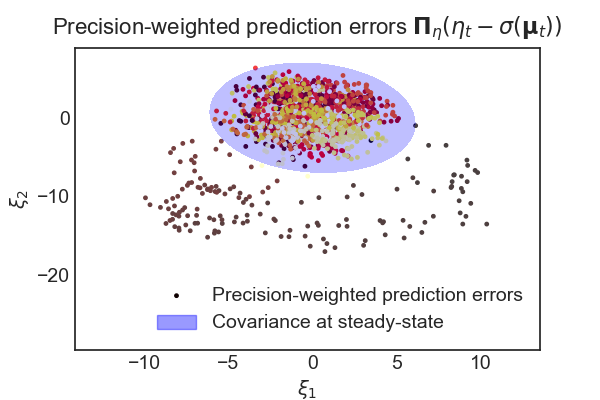}
    \caption{\textbf{Variational inference and predictive processing, averaging internal variables for any blanket state}. This figure illustrates a system's behaviour after experiencing a surprising blanket state, averaging internal variables for any blanket state. This is a multidimensional Ornstein-Uhlenbeck process, with two external, blanket and internal variables, initialised at the steady-state density conditioned upon an improbable blanket state $p(x_0 |b_0)$. \textit{Upper left:} we plot a sample trajectory of the blanket states as these relax to steady-state over a contour plot of the free energy (up to a constant). \textit{Upper right:} this plots the free energy (up to a constant) over time, averaged over multiple trajectories. In this example, the rare fluctuations that climb the free energy landscape vanish on average, so that the average free energy decreases monotonically. This need not always be the case: conservative systems (i.e., $\varsigma \equiv 0$ in \eqref{eq: Gaussian preserving diffusions}) are deterministic flows along the contours of the steady-state density (see Appendix \ref{app: helmholtz decomposition}). Since these contours do not generally coincide with those of $F(b , \boldsymbol{\mu})$ it follows that the free energy oscillates between its maximum and minimum value over the system's periodic trajectory. Luckily, conservative systems are not representative of dissipative, living systems. Yet, it follows that the average free energy of expected internal variables may increase, albeit only momentarily, in dissipative systems \eqref{eq: Gaussian preserving diffusions} whose solenoidal flow dominates dissipative flow. \textit{Lower left}: we illustrate the accuracy of predictions over external states of the sample path from the upper left panel. At steady-state (from timestep $\sim 100$), the predictions become accurate. The prediction of the second component is offset by four units for greater visibility, as can be seen from the longtime behaviour converging to four instead of zero. \textit{Lower right:} We show the evolution of precision-weighted prediction errors $\xi_t := \boldsymbol{\Pi}_{\eta}(\eta_t - \sigma(\boldsymbol{\mu}_t))$ over time. These are normally distributed with zero mean at steady-state.}
    \label{fig: Bayesian mechanics 6d}
\end{figure}

\begin{figure}
    \centering
    \includegraphics[width= 0.47\textwidth]{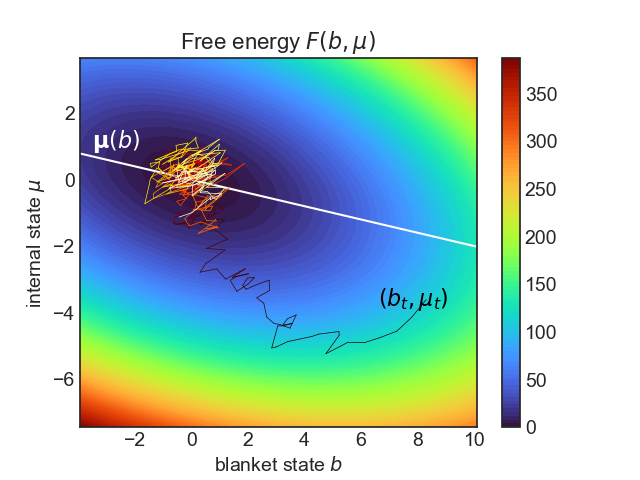}
    \includegraphics[width= 0.47\textwidth]{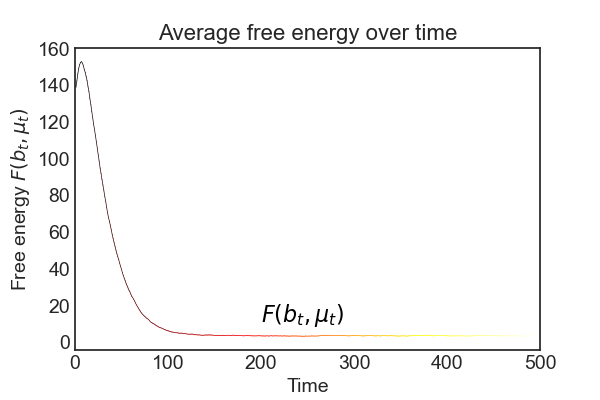}
    \includegraphics[width= 0.47\textwidth]{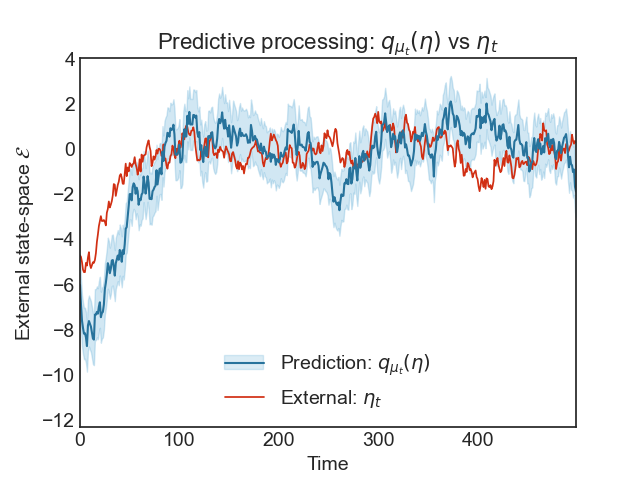}
    \includegraphics[width= 0.47\textwidth]{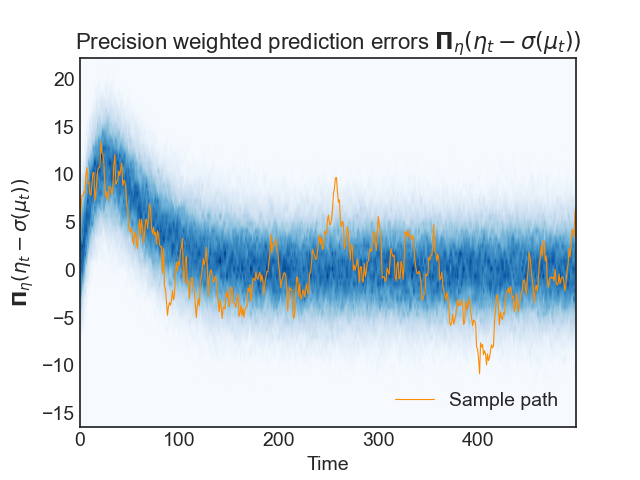}
    \caption{\textbf{Variational inference and predictive processing}. This figure illustrates a system's behaviour after experiencing a surprising blanket state. This is a multidimensional Ornstein-Uhlenbeck process, with one external, blanket and internal variable, initialised at the steady-state density conditioned upon an improbable blanket state $p(x_0 |b_0)$.
    \textit{Upper left:} this plots a sample trajectory of particular states as these relax to steady-state over a contour plot of the free energy. The white line shows the expected internal state given blanket states, at which point inference is exact. After starting close to this line, the process is driven by solenoidal flow to regions where inference is inaccurate. Yet, solenoidal flow makes the system converge faster to steady-state \cite{ottobreMarkovChainMonte2016,rey-belletIrreversibleLangevinSamplers2015} at which point inference becomes accurate again. \textit{Upper right:} this plots the free energy (up to a constant) over time, averaged over multiple trajectories. \textit{Lower left:} we illustrate the accuracy of predictions over external states of the sample path from the upper left panel. These predictions are accurate at steady-state (from timestep $\sim 100$). \textit{Lower right:} we illustrate the (precision weighted) prediction errors over time. In orange we plot the prediction error corresponding to the sample path in the upper left panel; the other sample paths are summarised as a heat map in blue.}
    \label{fig: Bayesian mechanics 3d}
\end{figure}

We can go further and associate expected internal states to the solution to the classical variational inference problem from statistical machine learning \cite{bleiVariationalInferenceReview2017} and theoretical neurobiology \cite{bogaczTutorialFreeenergyFramework2017,buckleyFreeEnergyPrinciple2017}. Expected internal states are the unique minimiser of a free energy functional (i.e., an evidence bound \cite{bealVariationalAlgorithmsApproximate2003,bleiVariationalInferenceReview2017})
    \begin{equation}
    \label{eq: free energy def}
    \begin{split}
        F(b_t, \mu_t) &\geq F(b_t, \boldsymbol \mu_t)\\
        F(b, \mu) &= \dkl[q_{\mu}(\eta)\| p(\eta|b)] -\log p(b, \mu)\\
        &= \underbrace{\E_{q_{\mu}(\eta)}[-\log p(x) ]}_{\text{Energy}}- \underbrace{\H[q_\mu]}_{\text{Entropy}}.
    \end{split}
    \end{equation}
The last line expresses the free energy as a difference between energy and entropy: energy or accuracy measures to what extent predicted external states are close to the true external states, while entropy penalises beliefs that are overly precise.

At first sight, variational inference and predictive processing are solely useful to characterise the average internal state given blanket states at steady-state. It is then surprising to see that the free energy says a great deal about a system's expected trajectories as it relaxes to steady-state. Figure \ref{fig: Bayesian mechanics 6d} and \ref{fig: Bayesian mechanics 3d} illustrate the time-evolution of the free energy and prediction errors after exposure to a surprising stimulus. In particular, Figure \ref{fig: Bayesian mechanics 6d} averages internal variables for any blanket state: In the neurosciences, perhaps the closest analogy is the event-triggered averaging protocol, where neurophysiological responses are averaged following a fixed perturbation, such a predictable neural input or an experimentally-controlled sensory stimulus (e.g., spike-triggered averaging, event-related potentials) \cite{schwartzSpiketriggeredNeuralCharacterization2006,sayerTimeCourseAmplitude1990,luckIntroductionEventRelatedPotential2014}.

The most striking observation is the nearly monotonic decrease of the free energy as the system relaxes to steady-state. This simply follows from the fact that regions of high density under the steady-state distribution have a low free energy. This \textit{overall decrease} in free energy is the essence of the free-energy principle, which describes self-organisation at non-equilibrium steady-state \cite{fristonFreeenergyPrincipleUnified2010,fristonFreeEnergyPrinciple2019a,parrMarkovBlanketsInformation2020}. Note that the free energy, even after averaging internal variables, may decrease non-monotonically. See the explanation in Figure \ref{fig: Bayesian mechanics 6d}.

\section{Active inference and stochastic control}
\label{sec: active inference and stoch control}

\begin{figure}[!h]
\centering\includegraphics[width=0.7\textwidth]{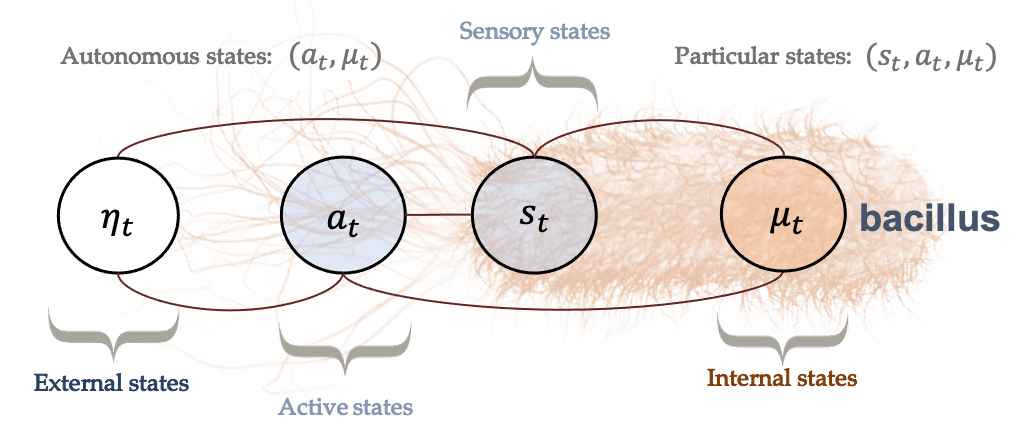}
\caption{\textbf{Markov blanket evolving in time comprising sensory and active states.} We continue the intuitive example from Figure \ref{fig: 3 way dynamical MB} of the bacillus as representing a Markov blanket that persists over time. The only difference is that we partition blanket states into sensory and active states. In this example, the sensory states can be seen as the bacillus' membrane, while the active states correspond to the actin filaments of the cytoskeleton.}
\label{fig: 4 way dynamical MB}
\end{figure}

In order to model agents that interact with their environment, we now partition blanket states into sensory and active states

    \begin{align*}
        b_t &= (s_t,a_t)\\
        x_t &= (\eta_t, s_t, a_t, \mu_t).
    \end{align*}
Intuitively, sensory states are the sensory receptors of the system (e.g., olfactory or visual receptors) while active states correspond to actuators through which the system influences the environment (e.g., muscles). See Figure \ref{fig: 4 way dynamical MB}. The goal of this section is to explain how autonomous states (i.e., active and internal states) respond adaptively to sensory perturbations in order to maintain the steady-state, which we interpret as the agent's preferences or goal. This allows us to relate the dynamics of autonomous states to active inference and stochastic control, which are well-known formulations of adaptive behaviour in theoretical biology and engineering.

\subsection{Active inference}

We now proceed to characterise autonomous states, given sensory states, using the free energy. Unpacking blanket states, the free energy \eqref{eq: free energy def} reads
    \begin{align*}
    F(s, a, \mu) &=\dkl[q_{\mu}(\eta)\| p(\eta|s,a)]-\log p( \mu |s,a ) -\log p(a|s ) -\log p(s ).
    \end{align*}

\begin{figure}
    \centering
    \includegraphics[width= 0.47\textwidth]{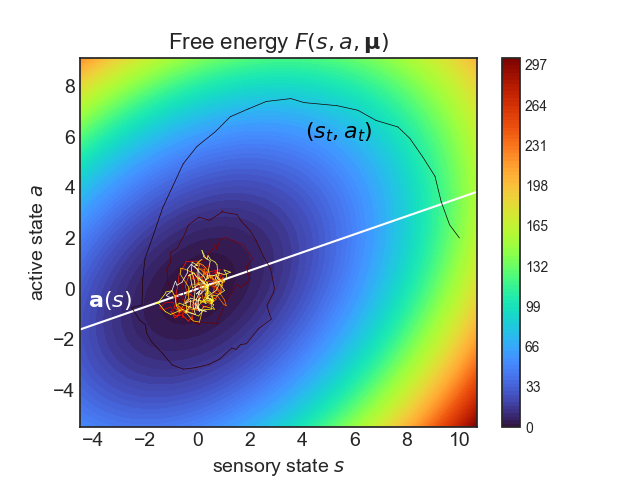}
    \includegraphics[width= 0.47\textwidth]{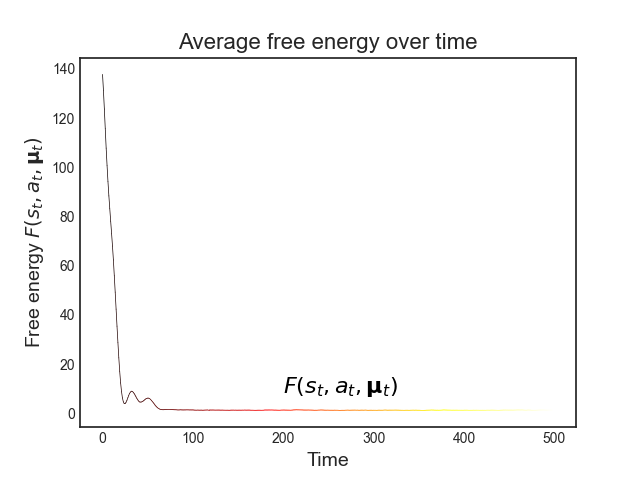}
    \caption{\textbf{Active inference}. 
    This figure illustrates a system's behaviour after experiencing a surprising sensory state, averaging internal variables for any blanket state. We simulated an Ornstein-Uhlenbeck process with two external, one sensory, one active and two internal variables, initialised at the steady-state density conditioned upon an improbable sensory state $p(x_0 |s_0)$. \textit{Left:} The white line shows the expected active state given sensory states: this is the action that performs active inference and optimal stochastic control. As the process experiences a surprising sensory state, it initially relaxes to steady-state in a winding manner due to the presence of solenoidal flow. Even though solenoidal flow drives the actions away from the optimal action initially, it allows the process to converge faster to steady-state \cite{rey-belletIrreversibleLangevinSamplers2015,ottobreMarkovChainMonte2016,lelievreOptimalNonreversibleLinear2013} where the actions are again close to the optimal action from optimal control. \textit{Right:} We plot the free energy of the expected internal state, averaged over multiple trajectories. In this example, the average free energy does not decrease monotonically---see Figure \ref{fig: Bayesian mechanics 6d} for an explanation.}
    \label{fig: active inference}
\end{figure}

Crucially, it follows that the expected autonomous states minimise free energy
    \begin{align*}
        F(s_t, a_t, \mu_t) &\geq F( s_t, \boldsymbol a_t,\boldsymbol \mu_t),\\
    \boldsymbol a_t := \boldsymbol a(s_t) &:=  \E_{p(a_t|s_t)}[a_t] =\Sigma_{a s} \Sigma_{s}^{-1} s_t,
    \end{align*}
where $\boldsymbol a_t$ denotes the expected active states given sensory states, which is the mean of $p(a_t|s_t)$. This result forms the basis of active inference, a well-known framework to describe and generate adaptive behaviour in neuroscience, machine learning and robotics \cite{buckleyFreeEnergyPrinciple2017,ueltzhofferDeepActiveInference2018,millidgeDeepActiveInference2020,heinsDeepActiveInference2020,lanillosRobotSelfOther2020,verbelenActiveInferenceFirst2020,adamsPredictionsNotCommands2013,pezzatoNovelAdaptiveController2020,oliverEmpiricalStudyActive2021,fristonActionBehaviorFreeenergy2010}. See Figure \ref{fig: active inference}.

\subsection{Multivariate control}

Active inference is used in various domains to simulate control \cite{koudahlWorkedExampleFokkerPlanckBased2020,verbelenActiveInferenceFirst2020,oliverEmpiricalStudyActive2021,ueltzhofferDeepActiveInference2018,fristonWhatOptimalMotor2011,sancaktarEndtoEndPixelBasedDeep2020,baltieriPIDControlProcess2019,pezzatoNovelAdaptiveController2020}, thus, it is natural that we can relate the dynamics of active states to well-known forms of stochastic control.

\begin{figure}
    \centering
    \includegraphics[width= 0.47\textwidth]{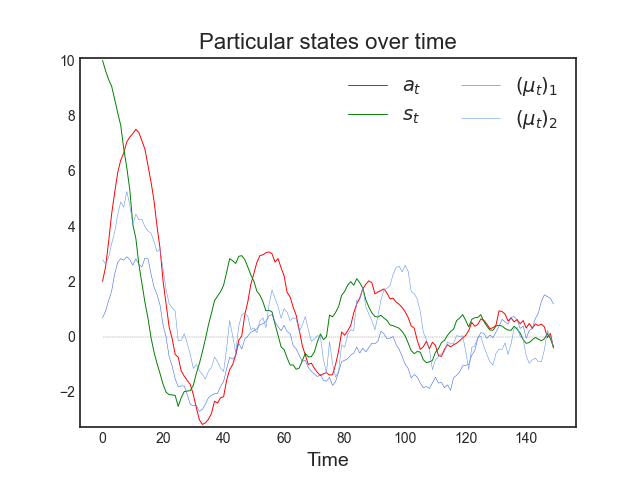}
    \caption{\textbf{Stochastic control}. This figure plots a sample path of the system's particular states after it experiences a surprising sensory state. This is the same sample path as shown in Figure 8 (left panel), however, here the link with stochastic control is easier to see. Indeed, it looks as if active states (in red) are actively compensating for sensory states (in green): rises in the active state-space lead to plunges in the sensory state-space and vice-versa. Notice the initial rise in active states to compensate for the perturbation in the sensory states. Active states follow a similar trajectory as sensory states, with a slight delay, which can be interpreted as a reaction time \cite{kosinski2008literature}. In fact, the correspondence between sensory and active states is a consequence of the solenoidal flow--see Figure 8 (left panel). The damped oscillations as the particular states approach their target value of $0$ (in grey) is analogous to that found in basic implementations of stochastic control, e.g., \cite[Figure 4.9]{roskillyMarineSystemsIdentification2015}.}
    \label{fig: stochastic control}
\end{figure}

By computing the free energy explicitly (see Appendix \ref{app: free energy}), we obtain that

\begin{align}
   (\boldsymbol a_t, \boldsymbol \mu_t) \quad &\text{minimises}\quad  (a, \mu) \mapsto \begin{bmatrix}s_t,a, \mu  \end{bmatrix} K \begin{bmatrix}s_t\\a\\ \mu  \end{bmatrix} \label{eq: multivariate control}\\
   K &:= \Sigma_{b:\mu}^{-1} \nonumber
\end{align}
where we denoted by $K$ the concentration (i.e., precision) matrix of $p(s,a, \mu)$. We may interpret $(\boldsymbol a, \boldsymbol \mu)$ as controlling how far particular states $\left[s,a, \mu  \right]$ are from their target set-point of $\left[0,0, 0  \right]$, where the error is weighted by the precision matrix $K$. See Figure \ref{fig: stochastic control}. (Note that we could choose any other set-point by translating the frame of reference or equivalently choosing a Gaussian steady-state centred away from zero). In other words, there is a cost associated to how far away $s, a, \mu$ are from the origin and this cost is weighed by the precision matrix, which derives from the stationary covariance of the steady-state. In summary, the expected internal and active states can be seen as performing multivariate stochastic control, where the matrix $K$ encodes control gains. From a biologist’s perspective, this corresponds to a simple instance of homeostatic regulation: maintaining physiological variables within their preferred range.

\subsection{Stochastic control in an extended state-space}

More sophisticated control methods, such as PID (Proportional-Integral-Derivative) control \cite{astromPidControllers1995,baltieriPIDControlProcess2019}, involve controlling a process and its higher orders of motion (e.g., integral or derivative terms). So how can we relate the dynamics of autonomous states to these more sophisticated control methods? The basic idea involves extending the sensory state-space to replace the sensory process $s_t$ by its various orders of motion $\tilde s_t = \left(s^{(0)}_t, \ldots, s^{(n)}_t \right)$ (integral, position, velocity, jerk etc, up to order $n$). To find these orders of motion, one must solve the stochastic realisation problem.

\subsubsection{The stochastic realisation problem}
\label{sec: stochastic realisation}

Recall that the sensory process $s_t$ is a stationary stochastic process (with a Gaussian steady-state). The following is a central problem in stochastic systems theory: Given a stationary stochastic process $s_t$, find a Markov process $\tilde s_t$, called the state process, and a function $f$ such that 
\begin{align}
\label{eq: stochastic realisation problem}
    s_t=f(\tilde s_t) \quad \text{for all}\quad  t.
\end{align}
Moreover, find an Itô stochastic differential equation whose unique solution is the state process $\tilde s_t$. The problem of characterising the family of all such representations is known as the stochastic realisation problem \cite{mitterTheoryNonlinearStochastic1981}.

What kind of processes $s_t$ can be expressed as a function of a Markov process \eqref{eq: stochastic realisation problem}?

There is a rather comprehensive theory of stochastic realisation for the case where $s_t$ is a Gaussian process (which occurs, for example, when $x_t$ is a Gaussian process). This theory expresses $s_t$ as a linear map of an Ornstein-Uhlenbeck process \cite{lindquistLinearStochasticSystems2015,lindquistRealizationTheoryMultivariate1985,pavliotisStochasticProcessesApplications2014}. The idea is as follows: as a mean-zero Gaussian process, $s_t$ is completely determined by its autocovariance function $C(t-r)=\E\left[s_{t} \otimes s_{r}\right]$, which by stationarity only depends on $|t-r|$. It is well known that any mean-zero stationary Gaussian process with exponentially decaying autocovariance function is an Ornstein-Uhlenbeck process (a result sometimes known as Doob's theorem) \cite{doobBrownianMovementStochastic1942,wangTheoryBrownianMotion2014,rey-belletOpenClassicalSystems2006,pavliotisStochasticProcessesApplications2014}. Thus if $C$ equals a finite sum of exponentially decaying functions, we can express $s_t$ as a linear function of several nested Ornstein-Uhlenbeck processes, i.e., as an integrator chain from control theory \cite{kryachkovFinitetimeStabilizationIntegrator2010,zimenkoFinitetimeFixedtimeStabilization2018}

\begin{equation}
\label{eq: integrator chain}
\begin{split}
    s_t &= f(s^{(0)}_t)\\
    ds^{(0)}_t &= f_0(s^{(0)}_t, s^{(1)}_t) dt + \varsigma_0 dW^{(0)}_t \\
    ds^{(1)}_t &= f_1(s^{(1)}_t, s^{(2)}_t) dt + \varsigma_1 dW^{(1)}_t \\
    \vdots\quad  & \qquad \vdots\qquad \vdots \\
    ds^{(n-1)}_t &= f_{n-1}(s^{(n-1)}_t, s^{(n)}_t) dt + \varsigma_{n-1} dW^{(n-1)}_t \\
    ds^{(n)}_t &= f_n(s^{(n)}_t) dt + \varsigma_n dW^{(n)}_t. \\
\end{split}
\end{equation}

\begin{figure}
    \centering
    \includegraphics[width=0.5\textwidth]{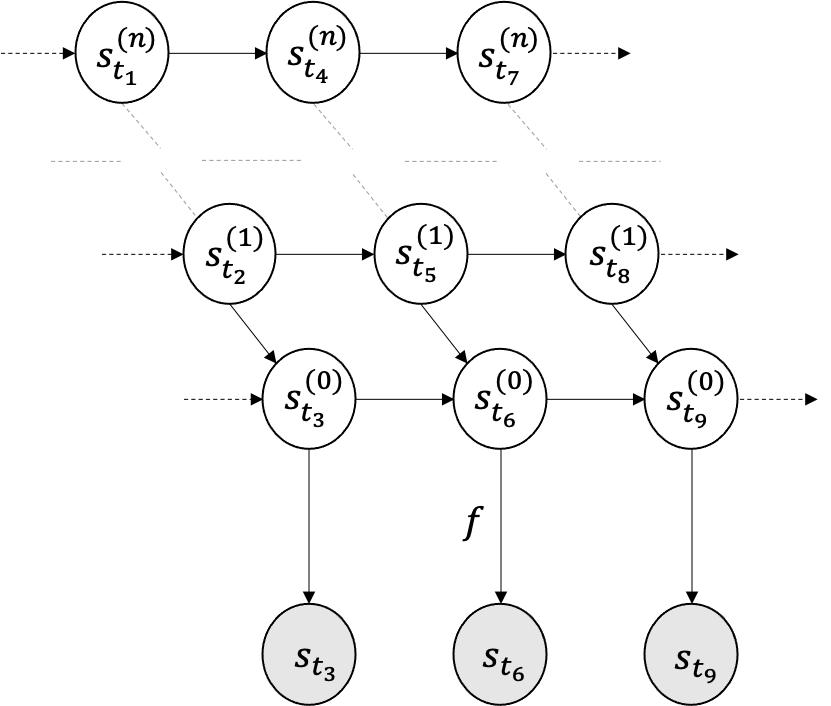}
    \caption{\textbf{Continuous-time Hidden Markov model}. This figure depicts \eqref{eq: integrator chain} in a graphical format, as a Bayesian network \cite{wainwrightGraphicalModelsExponential2007,pearlGraphicalModelsProbabilistic1998}. The encircled variables are random variables---the processes indexed at an arbitrary sequence of subsequent times $t_1<t_2<\ldots <t_9$. The arrows represent relationships of causality. In this hidden Markov model, the (hidden) state process $\tilde s_t$ is given by an integrator chain---i.e., nested stochastic differential equations $s^{(0)}_t, s^{(1)}_t, \ldots, s^{(n)}_t$. These processes $s^{(i)}_t, i\geq 0$, can respectively be seen as encoding the position, velocity, jerk etc, of the process $s_t$.}
    \label{fig: HMM}
\end{figure}

In this example, $f,f_i$ are suitably chosen linear functions, $\varsigma_i$ are matrices and $W^{(i)}$ are standard Brownian motions. Thus, we can see $s_t$ as the output of a continuous-time hidden Markov model, whose (hidden) states $s^{(i)}_t$ encode its various orders of motion: position, velocity, jerk etc. These are known as generalised coordinates of motion in the Bayesian filtering literature \cite{fristonVariationalFiltering2008,fristonVariationalTreatmentDynamic2008,fristonGeneralisedFiltering2010}. See Figure \ref{fig: HMM}. 

More generally, the state process $\tilde s_t$ and the function $f$ need not be linear, which enables to realise non-linear, non-Gaussian processes $s_t$ \cite{fristonVariationalFiltering2008, parrComputationalNeurologyMovement2021,gomesMeanFieldLimits2020}. Technically, this follows as Ornstein-Uhlenbeck processes are the only stationary Gaussian Markov processes. Note that stochastic realisation theory is not as well developed in this general case \cite{tayorNonlinearStochasticRealization1989,mitterTheoryNonlinearStochastic1981,frazhoStochasticRealizationTheory1982,gomesMeanFieldLimits2020,fristonVariationalFiltering2008}.

\subsubsection{Stochastic control of integrator chains}

Henceforth, we assume that we can express $s_t$ as a function of a Markov process $\tilde s_t$ \eqref{eq: stochastic realisation problem}. Inserting \eqref{eq: stochastic realisation problem} into \eqref{eq: multivariate control}, we now see that the expected autonomous states minimise how far themselves and $f(\tilde s_t)$ are from their target value of zero

\begin{align}
\label{eq: stoch control of integrator chains}
 (\boldsymbol a_t, \boldsymbol \mu_t) \quad \text{minimises}\quad (a, \mu)\mapsto  \begin{bmatrix}f(\tilde s_t),a, \mu  \end{bmatrix}  K \begin{bmatrix}f(\tilde s_t)\\a\\ \mu  \end{bmatrix}.
\end{align}

Furthermore, if the state process $\tilde s_t$ can be expressed as an integrator chain, as in \eqref{eq: integrator chain}, then we can interpret expected active and internal states as controlling each order of motion $s^{(i)}_t$. For example, if $f$ is linear, these processes control each order of motion $s^{(i)}_t$ towards its target value of zero.

\subsubsection{PID-like control}

Proportional-integral-derivative (PID) control is a well-known control method in engineering \cite{astromPidControllers1995,baltieriPIDControlProcess2019}. More than 90\% of controllers in engineered systems implement either PID or PI (no derivative) control. The goal of PID control is to control a signal $s^{(1)}_t$, its integral $s^{(0)}_t$, and its derivative $s^{(2)}_t$ close to a pre-specified target value \cite{baltieriPIDControlProcess2019}.

This turns out to be exactly what happens here when we consider the stochastic control of an integrator chain \eqref{eq: stoch control of integrator chains} with three orders of motion $(n=2)$. When $f$ is linear, expected autonomous states control integral, proportional and derivative processes $s^{(0)}_t,s^{(1)}_t,s^{(2)}_t$ towards their target value of zero. Furthermore, from $f$ and $K$ one can derive integral, proportional and derivative gains, which penalise deviations of $s^{(0)}_t,s^{(1)}_t,s^{(2)}_t$, respectively, from their target value of zero. Crucially, these control gains are simple by-products of the steady-state density and the stochastic realisation problem.

Why restrict ourselves to PID control when stochastic control of integrator chains is available? It turns out that when sensory states $s_t$ are expressed as a function of an integrator chain \eqref{eq: integrator chain}, one may get away by controlling an approximation of the true (sensory) process, obtained by truncating high orders of motion as these have less effect on the dynamics, though knowing when this is warranted is a problem in approximation theory. This may explain why integral feedback control ($n=0$), PI control ($n=1$) and PID control ($n=2$) are the most ubiquitous control methods in engineering applications. However, when simulating biological control---usually with highly non-linear dynamics---it is not uncommon to consider generalised motion to fourth ($n=4$) or sixth ($n=6$) order \cite{fristonGraphicalBrainBelief2017,parrComputationalNeurologyMovement2021}.

It is worth mentioning that PID control has been shown to be implemented in simple molecular systems and is becoming a popular mechanistic explanation of behaviours such as bacterial chemotaxis and robust homeostatic algorithms in biochemical networks \cite{baltieriPIDControlProcess2019,chevalierDesignAnalysisProportionalIntegralDerivative2019,yiRobustPerfectAdaptation2000}. We suggest that this kind of behaviour emerges in Markov blankets at non-equilibrium steady-state. Indeed, stationarity means that autonomous states will look as if they respond adaptively to external perturbations to preserve the steady-state, and we can identify these dynamics as implementations of various forms of stochastic control (including PID-like control).

\section{Discussion}

In this paper, we considered the consequences of a boundary mediating interactions between states internal and external to a system. On unpacking this notion, we found that the states internal to a Markov blanket look as if they perform variational Bayesian inference, optimising beliefs about their external counterparts. When subdividing the blanket into sensory and active states, we found that autonomous states perform active inference and various forms of stochastic control (i.e., generalisations of PID control). 

\textbf{Interacting Markov blankets:}
The sort of inference we have described could be nuanced by partitioning the external state-space into several systems that are themselves Markov blankets (such as Markov blankets nested at several different scales \cite{hespMultiscaleViewEmergent2019}). From the perspective of internal states, this leads to a more interesting inference problem, with a more complex generative model. It may be that the distinction between the sorts of systems we generally think of as engaging in cognitive, inferential, dynamics \cite{fristonHierarchicalModelsBrain2008} and simpler systems rest upon the level of structure of the generative models (i.e., steady-state densities) that describe their inferential dynamics.

\textbf{Temporally deep inference:} This distinction may speak to a straightforward extension of the treatment on offer, from simply inferring an external state to inferring the trajectories of external states. This may be achieved by representing the external process in terms of its higher orders of motion by solving the stochastic realisation problem. By repeating the analysis above, internal states may be seen as inferring the position, velocity, jerk, etc of the external process, consistently with temporally deep inference in the sense of a Bayesian filter \cite{fristonGeneralisedFiltering2010} (a special case of which is an extended Kalman–Bucy filter \cite{kalmanNewApproachLinear1960}).

\textbf{Bayesian mechanics in non-Gaussian steady-states:} The treatment from this paper extends easily to non-Gaussian steady-states, in which internal states appear to perform approximate Bayesian inference over external states. Indeed, any arbitrary (smooth) steady-state density may be approximated by a Gaussian density at one of its modes using a so-called Laplace approximation. This Gaussian density affords one with a synchronisation map in closed form\footnote{Another option is to empirically fit a synchronisation map to data \cite{fristonLifeWeKnow2013}.} that maps the expected internal state to an approximation of the expected external state. It follows that the system can be seen as performing approximate Bayesian inference over external states---precisely, an inferential scheme known as variational Laplace \cite{fristonVariationalFreeEnergy2007}. We refer the interested reader to a worked-out example involving two sparsely coupled Lorenz systems \cite{fristonStochasticChaosMarkov2021}. Note that variational Laplace has been proposed as an implementation of various cognitive processes in biological systems \cite{buckleyFreeEnergyPrinciple2017,bogaczTutorialFreeenergyFramework2017, fristonActionBehaviorFreeenergy2010} accounting for several features of the brain’s functional anatomy and neural message passing \cite{fristonTheoryCorticalResponses2005,fristonHierarchicalModelsBrain2008,fristonPredictiveCodingFreeenergy2009,adamsPredictionsNotCommands2013,pezzuloActiveInferenceView2012}.

\textbf{Modelling real systems:} The simulations presented here are as simple as possible and are intended to illustrate general principles that apply to all stationary processes with a Markov blanket \eqref{eq: MB over time}. These principles have been used to account for synthetic data arising in more refined (and more specific) simulations of an interacting particle system \cite{fristonLifeWeKnow2013} and synchronisation between two sparsely coupled stochastic Lorenz systems \cite{fristonStochasticChaosMarkov2021}. Clearly, an outstanding challenge is to account for empirical data arising from more interesting and complex structures. To do this, one would have to collect time-series from an organism's internal states (e.g., neural activity), its surrounding external states, and its interface, including sensory receptors and actuators. Then, one could test for conditional independence between internal, external and blanket states \eqref{eq: MB over time} \cite{pelletUsingMarkovBlankets2008}. One might then test for the existence of a synchronisation map (using Lemma \ref{lemma: equiv sigma well defined}). This speaks to modelling systemic dynamics using stochastic processes with a Markov blanket. For example, one could learn the volatility, solenoidal flow and steady-state density in a stochastic differential equation \eqref{eq: Gaussian preserving diffusions} from data, using supervised learning \cite{tzenNeuralStochasticDifferential2019}.

\section{Conclusion}

This paper outlines some of the key relationships between stationary processes, inference and control. These relationships rest upon partitioning the world into those things that are internal or external to a (statistical) boundary, known as a Markov blanket. When equipped with dynamics, the expected internal states appear to engage in variational inference, while the expected active states appear to be performing active inference and various forms of stochastic control.

The rationale behind these findings is rather simple: if a Markov blanket derives from a steady-state density, the states of the system will look as if they are responding adaptively to external perturbations in order to recover the steady-state. Conversely, well-known methods used to build adaptive systems implement the same kind of dynamics, implicitly so that the system maintains a steady-state with its environment.

\vskip6pt
\enlargethispage{20pt}

\dataccess{All data and numerical simulations can be reproduced with code freely available at \url{https://github.com/conorheins/bayesian-mechanics-sdes}.}

\aucontribute{Conceptualization: LD, KF, CH, GAP; Formal analysis: LD, KF, GAP; Software: LD, CH; Supervision: KF, GAP; Writing – original draft: LD; Writing – review \& editing: KF, CH, GAP. All authors gave final approval for publication and agree to be held accountable for the work performed therein.}

\competing{We have no competing interests.}

\funding{LD is supported by the Fonds National de la Recherche, Luxembourg (Project code: 13568875). This publication is based on work partially supported by the EPSRC Centre for Doctoral Training in Mathematics of Random Systems: Analysis, Modelling and Simulation (EP/S023925/1). KF was a Wellcome Principal Research Fellow (Ref: 088130/Z/09/Z). CH is supported by the U.S. Office of Naval Research (N00014-19-1-2556). The work of GAP was partially funded by the EPSRC, grant number EP/P031587/1, and by JPMorgan Chase \& Co. Any views or opinions expressed herein are solely those of the authors listed, and may differ from the views and opinions expressed by JPMorgan Chase \& Co. or its affiliates. This material is not a product of the Research Department of J.P. Morgan Securities LLC. This material does not constitute a solicitation or offer in any jurisdiction.}

\ack{LD would like to thank Kai Ueltzhöffer, Toby St Clere Smithe and Thomas Parr for interesting discussions. We are grateful to our two anonymous reviewers for feedback which substantially improved the manuscript.}


\appendix

\section{Existence of synchronisation map: proof}
\label{app: sync map existence proof}

We prove Lemma \ref{lemma: equiv sigma well defined}.

\begin{proof}
$(i) \iff (ii)$ follows by definition of a function.

$(ii) \iff (iii)$ is as follows
\begin{equation*}
\begin{split}
   & \quad \forall b_1, b_2 \in \B: \boldsymbol \mu(b_1)= \boldsymbol \mu(b_2) \Rightarrow \boldsymbol \eta(b_1)= \boldsymbol \eta(b_2)\\
   &\iff \left(\forall b_1, b_2 \in \B: \Sigma_{\mu b} \Sigma_{b}^{-1} b_1 = \Sigma_{\mu b} \Sigma_{b}^{-1} b_2 \Rightarrow \Sigma_{\eta b} \Sigma_{b}^{-1} b_1 = \Sigma_{\eta b} \Sigma_{b}^{-1} b_2\right) \\
    &\iff \left(\forall b \in \B:  \Sigma_{\mu b} \Sigma_{b}^{-1} b=0 \Rightarrow \Sigma_{\eta b} \Sigma_{b}^{-1} b =0 \right) \\
    &\iff  \ker \Sigma_{\mu b} \subset  \ker \Sigma_{\eta b}\\
\end{split}
\end{equation*}

$(iii) \iff (iv)$ From \cite[Section 0.7.3]{hornMatrixAnalysisSecond2012}, using the Markov blanket condition \eqref{eq: MB is sparsity in precision matrix}, we can verify that
\begin{align*}
   \Pi_{\mu} \Sigma_{\mu b} &=- \Pi_{\mu b} \Sigma_b \\
   \Pi_{\eta}\Sigma_{\eta b} &=-  \Pi_{\eta b} \Sigma_b.
\end{align*}
Since $ \Pi_{\mu}, \Pi_{\eta}, \Sigma_b$ are invertible, we deduce
\begin{equation*}
\begin{split}
& \quad \ker \Sigma_{\mu b} \subset  \ker \Sigma_{\eta b}\\
&\iff \ker \Pi_{\mu} \Sigma_{\mu b} \subset  \ker \Pi_{\eta}\Sigma_{\eta b} \\
&\iff \ker -\Pi_{\mu b} \Sigma_b \subset  \ker -\Pi_{\eta b} \Sigma_b \\
&\iff \ker \Pi_{\mu b} \subset  \ker \Pi_{\eta b}.
\end{split}
\end{equation*}
\end{proof}

\section{The Helmholtz decomposition}
\label{app: helmholtz decomposition}

We consider a diffusion process $x_t$ on $\mathbb{R}^{d}$ satisfying an Itô stochastic differential equation (SDE) \cite{pavliotisStochasticProcessesApplications2014,rogersDiffusionsMarkovProcesses2000,oksendalStochasticDifferentialEquations2003},
\begin{equation}
\label{eq: diffusion process}
d x_{t}=f\left(x_{t}\right) d t+\varsigma\left(x_{t}\right) d W_{t},
\end{equation}
where $W_t$ is an $m$-dimensional standard Brownian motion (a.k.a., Wiener process) \cite{rogersDiffusionsMarkovProcesses2000a, pavliotisStochasticProcessesApplications2014} and $f:\mathbb{R}^{d} \rightarrow \mathbb{R}^{d}, \varsigma: \mathbb{R}^{d} \rightarrow \mathbb{R}^{d \times m}$ are smooth functions satisfying for all $x,y \in \R^d$:
\begin{itemize}
    \item Bounded, linear growth condition: $|f(x)|+|\varsigma(x)| \leq K(1+|x|)$,
    \item Lipschitz condition: $|f(x)-f(y)|+|\varsigma(x)-\varsigma( y)| \leq K|x-y|$,
\end{itemize}
for some constant $K$ and $|\varsigma|=\sum_{ij}\left|\varsigma_{i j}\right|$. These are standard regularity conditions that ensure the existence and uniqueness of a solution to the SDE \eqref{eq: diffusion process} \cite[Theorem 5.2.1]{oksendalStochasticDifferentialEquations2003}.

We now recall an important result from the theory of stationary diffusion processes, known as the Helmholtz decomposition. It consists of splitting the dynamic into time-reversible (i.e., dissipative) and time-irreversible (i.e., conservative) components. The importance of this result in non-equilibrium thermodynamics was originally recognised by Graham in 1977 \cite{grahamCovariantFormulationNonequilibrium1977} and has been of great interest in the field since \cite{eyinkHydrodynamicsFluctuationsOutside1996,aoPotentialStochasticDifferential2004,qianDecompositionIrreversibleDiffusion2013,pavliotisStochasticProcessesApplications2014}. Furthermore, the Helmholtz decomposition is widely used in statistical machine learning to generate Monte-Carlo sampling schemes  \cite{maCompleteRecipeStochastic2015,barpUnifyingCanonicalDescription2021,chaudhariStochasticGradientDescent2018,lelievreOptimalNonreversibleLinear2013,daiLargeBatchTraining2020,pavliotisStochasticProcessesApplications2014}.

\begin{lemma}[Helmholtz decomposition]
For a diffusion process \eqref{eq: diffusion process} and a smooth probability density $p>0$, the following are equivalent:
\begin{enumerate}
    \item $p$ is a steady-state for $x_t$.
    \item We can write the drift as
    \begin{equation}
    \label{eq: Helmholtz decomposition of drift}
    \begin{split}
        f &= f_{\text{rev}}+ f_{\text{irrev}}\\
        f_{\text{rev}} &:= \Gamma \nabla \log p + \nabla \cdot \Gamma\\
        f_{\text{irrev}} &:= Q \nabla \log p + \nabla \cdot Q. 
    \end{split}
    \end{equation}
    where $\Gamma= \varsigma \varsigma^\top/2$ is the diffusion tensor and $Q= -Q^\top$ is a smooth antisymmetric matrix field. $\nabla \cdot $ denotes the divergence of a matrix field defined as $\left(\nabla \cdot Q \right)_i := \sum_{j} \frac{\partial}{\partial x_j} Q_{ij}$.
\end{enumerate}
Furthermore, $f_{\text{rev}}$ is invariant under time-reversal, while $f_{\text{irrev}}$ changes sign under time-reversal.
\end{lemma}

\begin{figure}[!h]
\centering\includegraphics[width=0.45\textwidth]{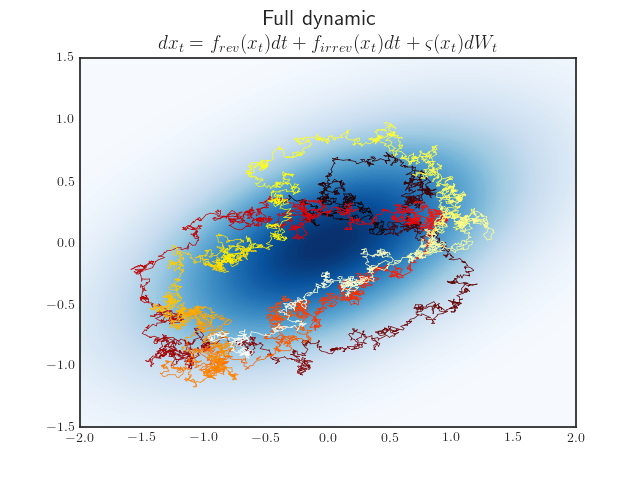}
\centering\includegraphics[width=0.45\textwidth]{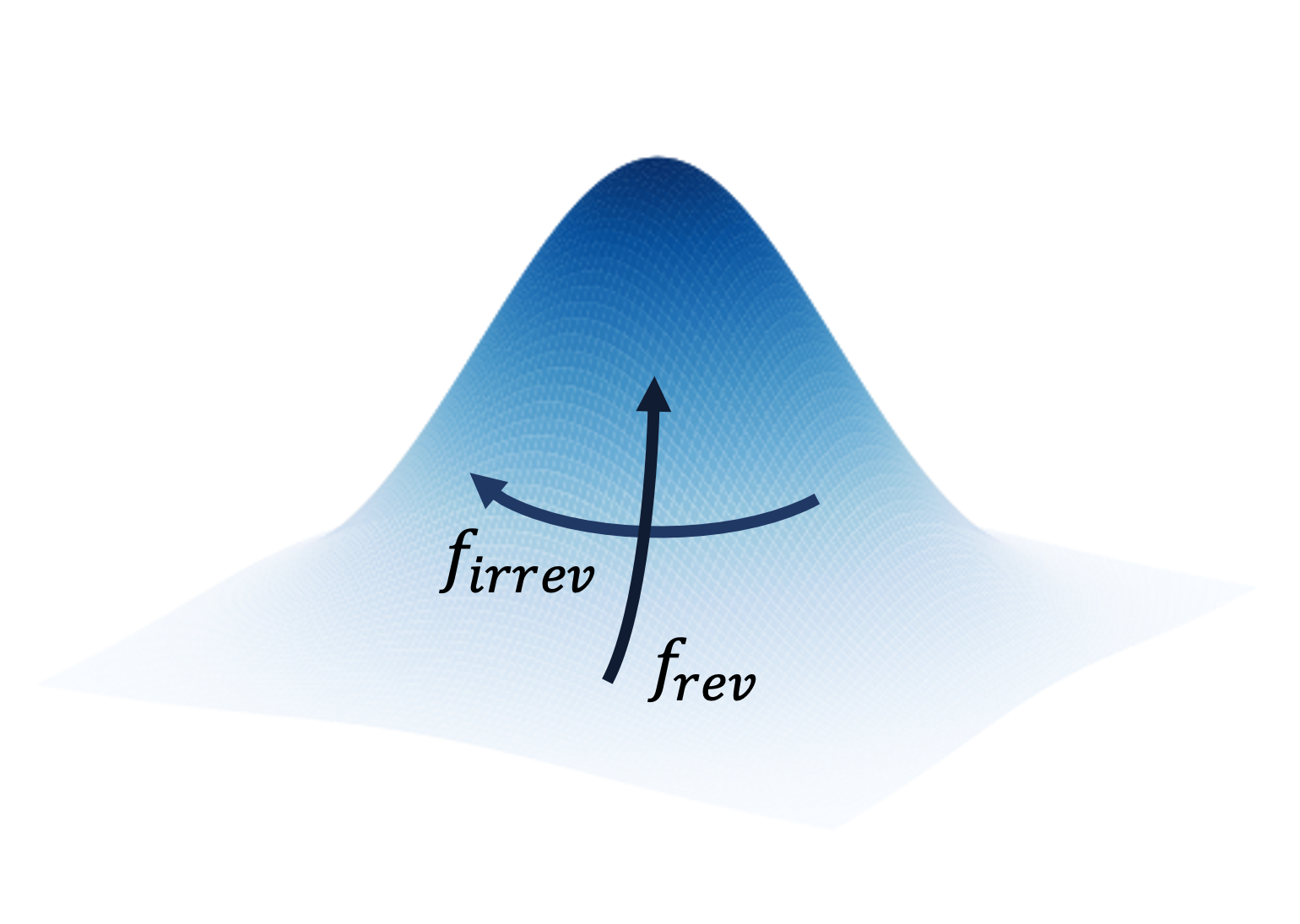}
\centering\includegraphics[width=0.45\textwidth]{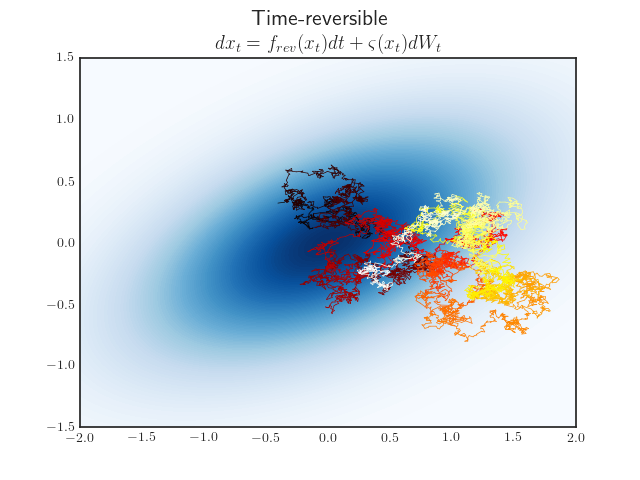}
\centering\includegraphics[width=0.45\textwidth]{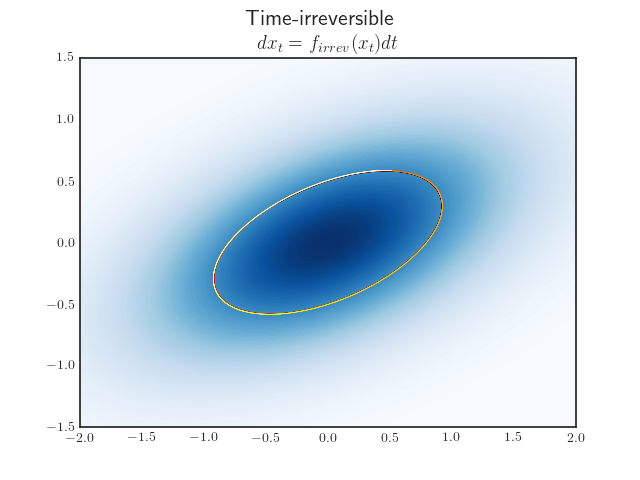}
\caption{\textbf{Helmholtz decomposition.} The upper left panel shows a sample trajectory of a two-dimensional diffusion process \eqref{eq: diffusion process} on a heat map of the (Gaussian) steady-state density. The upper right panel illustrates the Helmholtz decomposition of the drift into time-reversible and time-irreversible parts: the time-reversible part of the drift flows towards the peak of the steady-state density, while the time-irreversible part flows along the contours of the probability distribution. The lower panels plot sample paths of the time-reversible (lower left) and time-irreversible (lower right) parts of the dynamics. Purely conservative dynamics (lower right panel) are reminiscent of the trajectories of massive bodies (e.g., planets) whose random fluctuations are negligible, as in Newtonian mechanics. The lower panels help illustrate the meaning of time-irreversibility: If we were to reverse time (c.f., \eqref{eq: time reversed process}), the trajectories the time-reversible process would be, on average, no different, while the trajectories of the time-irreversible process would flow, say, clockwise instead of counterclockwise, which would clearly be distinguishable. Here, the full process (upper left panel) is a combination of both dynamics. As we can see the time-reversible part affords the stochasticity while the time-irreversible part characterises non-equilibria and the accompanying wandering behaviour that characterises life-like systems \cite{skinnerImprovedBoundsEntropy2021,aokiEntropyProductionLiving1995}.}
\label{fig: Helmholtz decomposition}
\end{figure}

In the Helmholtz decomposition of the drift \eqref{eq: Helmholtz decomposition of drift}, the diffusion tensor $\Gamma$ mediates the dissipative flow, which flows towards the modes of the steady-state density, but is counteracted by random fluctuations $W_t$, so that the system's distribution remains unchanged---together these form the time-reversible part of the dynamics. In contrast, $Q$ mediates the solenoidal flow---whose direction is reversed under time-reversal---which consists of conservative (i.e., Hamiltonian) dynamics that flow on the level sets of the steady-state. See Figure \ref{fig: Helmholtz decomposition} for an illustration. Note that the terms time-reversible and time-irreversible are meant in a probabilistic sense, in the sense that time-reversibility denotes invariance under time-reversal. This is opposite to reversible and irreversible in a classical physics sense, which respectively mean energy preserving (i.e., conservative) and entropy creating (i.e., dissipative).

\begin{proof}
\begin{itemize}
    \item["$\Rightarrow$"] It is well-known that when $x_t$ is stationary at $p$, its time-reversal is also a diffusion process that solves the following Itô SDE \cite{haussmannTimeReversalDiffusions1986}
    \begin{equation}
    \label{eq: time reversed process}
    \begin{split}
        dx_t^- &= f^-(x_t^-)dt + \varsigma(x_t^-) dW_t \\
        f^-&:=-b+p^{-1} \nabla \cdot\left(2\Gamma p\right).
    \end{split}
    \end{equation}
    This enables us to write the drift $f$ as a sum of two terms: one that is invariant under time reversal, another that changes sign under time-reversal
    \begin{equation*}
        f = \frac{f + f^-}{2} + \frac{f - f^-}{2} =: f_{\text{rev}}+ f_{\text{irrev}}.
    \end{equation*}
    It is straightforward to identify the time-reversible term
    \begin{align}
    \label{eq: time reversible drift}
        f_{\text{rev}} &= p^{-1}\nabla \cdot (\Gamma p) = p^{-1}\Gamma \nabla p +p^{-1}p \nabla \cdot \Gamma = \Gamma \nabla \log p + \nabla \cdot \Gamma.
    \end{align}
    For the remaining term, we first note that the steady-state $p$ solves the stationary Fokker-Planck equation \cite{pavliotisStochasticProcessesApplications2014,riskenFokkerPlanckEquationMethods1996}
    \begin{equation*}
        \nabla \cdot (-f p + \nabla \cdot (\Gamma p))=0.
    \end{equation*}
    Decomposing the drift into time-reversible and time-irreversible parts, we obtain that the time-irreversible part produces divergence-free (i.e., conservative) flow w.r.t. the steady-state distribution
    \begin{equation*}
        \nabla \cdot (f_{\text{irrev}}p)=0.
    \end{equation*}
    Now recall that any (smooth) divergence-free vector field is the divergence of a (smooth) antisymmetric matrix field $A=-A^T$ \cite{grahamCovariantFormulationNonequilibrium1977, eyinkHydrodynamicsFluctuationsOutside1996, RealAnalysisEvery}
    \begin{equation*}
        f_{\text{irrev}}p = \nabla \cdot A.
    \end{equation*}
    We define a new antisymmetric matrix field $Q:= p^{-1}A$. It follows from the product rule for divergences that we can rewrite the time-irreversible drift as required
    \begin{equation*}
       f_{\text{irrev}} = Q \nabla \log p + \nabla \cdot Q.
    \end{equation*}
    \item["$\Leftarrow$"] From \eqref{eq: time reversible drift} we can rewrite the time-reversible part of the drift as 
    \begin{align}
        f_{\text{rev}} &= p^{-1}\nabla \cdot (\Gamma p).
    \end{align}
    In addition, we define the auxiliary antisymmetric matrix field $A := pQ$ and use the product rule for divergences to simplify the expression of the time-irreversible part
    \begin{equation*}
        f_{\text{irrev}} = p^{-1}\nabla \cdot A.
    \end{equation*}
    Note that 
    \begin{equation*}
       \nabla \cdot( f_{\text{irrev}}p) = 0
    \end{equation*}
    as the matrix field $A$ is smooth and antisymmetric. It follows that the distribution $p$ solves the stationary Fokker-Planck equation
    \begin{align*}
        \nabla \cdot (- f p + \nabla \cdot (\Gamma p)) &= \nabla \cdot (-f_{\text{rev}}p-f_{\text{irrev}}p + \nabla \cdot (\Gamma p))= \nabla \cdot (-f_{\text{irrev}}p) =0.
    \end{align*}
\end{itemize}
\end{proof}

\section{Free energy computations}
\label{app: free energy}

The free energy reads \eqref{eq: free energy def}
\begin{align*}
    F(b, \mu) &=\dkl[q_{\mu}(\eta)\| p(\eta|b)]-\log p(b, \mu ).
\end{align*}
Recalling from \eqref{eq: posterior beliefs}, \eqref{eq: def approx posterior} that $q_{\mu}(\eta)$ and $ p(\eta|b)$ are Gaussian, the KL divergence between multivariate Gaussians is well-known
\begin{align*}
    &q_{\mu}(\eta) = \mathcal N(\eta ; \sigma(\mu), \Pi_\eta^{-1}), \quad p(\eta|b) = \mathcal N(\eta ; \boldsymbol \eta(b), \Pi_\eta^{-1}), \\
    &\Rightarrow \dkl[q_{\mu}(\eta)\| p(\eta|b)] = \frac 1 2 (\sigma(\mu)-\boldsymbol \eta(b))\Pi_\eta(\sigma(\mu)-\boldsymbol \eta(b)).
\end{align*}
Furthermore, we can compute the log partition
\begin{align*}
    -\log p(b, \mu) = \frac 1 2 \begin{bmatrix}b, \mu  \end{bmatrix} \Sigma_{b:\mu}^{-1} \begin{bmatrix}b \\ \mu  \end{bmatrix} \quad \text{(up to a constant)}.
\end{align*}
Note that $\Sigma_{b:\mu}^{-1}$ is the inverse of a principal submatrix of $\Sigma$, which in general differs from $\Pi_{b:\mu}$, a principal submatrix of $\Pi$. Finally,
\begin{align*}
    F(b, \mu) &= \frac 1 2 (\sigma(\mu)-\boldsymbol \eta(b))\Pi_\eta(\sigma(\mu)-\boldsymbol \eta(b)) +\frac 1 2 \begin{bmatrix}b, \mu  \end{bmatrix} \Sigma_{b:\mu}^{-1} \begin{bmatrix}b \\ \mu  \end{bmatrix}  \quad \text{(up to a constant)}.
\end{align*}


\bibliographystyle{unsrt}
\bibliography{bib}

\begin{thebibliography}{100}

\bibitem{hespMultiscaleViewEmergent2019}
Casper Hesp, Maxwell Ramstead, Axel Constant, Paul Badcock, Michael Kirchhoff,
  and Karl Friston.
\newblock A {{Multi}}-scale {{View}} of the {{Emergent Complexity}} of
  {{Life}}: {{A Free}}-{{Energy Proposal}}.
\newblock In Georgi~Yordanov Georgiev, John~M. Smart, Claudio~L.
  Flores~Martinez, and Michael~E. Price, editors, {\em Evolution,
  {{Development}} and {{Complexity}}}, Springer {{Proceedings}} in
  {{Complexity}}, pages 195--227, {Cham}, 2019. {Springer International
  Publishing}.

\bibitem{kirchhoffMarkovBlanketsLife2018}
Michael Kirchhoff, Thomas Parr, Ensor Palacios, Karl Friston, and Julian
  Kiverstein.
\newblock The {{Markov}} blankets of life: Autonomy, active inference and the
  free energy principle.
\newblock {\em Journal of The Royal Society Interface}, 15(138):20170792,
  January 2018.

\bibitem{pearlGraphicalModelsProbabilistic1998}
Judea Pearl.
\newblock Graphical {{Models}} for {{Probabilistic}} and {{Causal Reasoning}}.
\newblock In Philippe Smets, editor, {\em Quantified {{Representation}} of
  {{Uncertainty}} and {{Imprecision}}}, Handbook of {{Defeasible Reasoning}}
  and {{Uncertainty Management Systems}}, pages 367--389. {Springer
  Netherlands}, {Dordrecht}, 1998.

\bibitem{bishopPatternRecognitionMachine2006}
Christopher~M. Bishop.
\newblock {\em Pattern Recognition and Machine Learning}.
\newblock Information Science and Statistics. {Springer}, {New York}, 2006.

\bibitem{nicolisSelforganizationNonequilibriumSystems1977}
G.~Nicolis and I.~Prigogine.
\newblock {\em Self-Organization in {{Nonequilibrium Systems}}: {{From
  Dissipative Structures}} to {{Order Through Fluctuations}}}.
\newblock {Wiley-Blackwell}, {New York}, June 1977.

\bibitem{goldbeterDissipativeStructuresBiological2018}
Albert Goldbeter.
\newblock Dissipative structures in biological systems: Bistability,
  oscillations, spatial patterns and waves.
\newblock {\em Philosophical Transactions of the Royal Society A: Mathematical,
  Physical and Engineering Sciences}, 376(2124):20170376, July 2018.

\bibitem{hakenSynergeticsIntroductionNonequilibrium1978}
Hermann Haken.
\newblock {\em Synergetics: {{An Introduction Nonequilibrium Phase
  Transitions}} and {{Self}}-{{Organization}} in {{Physics}}, {{Chemistry}} and
  {{Biology}}}.
\newblock Springer {{Series}} in {{Synergetics}}. {Springer-Verlag}, {Berlin
  Heidelberg}, second edition, 1978.

\bibitem{perunovStatisticalPhysicsAdaptation2016}
Nikolay Perunov, Robert~A. Marsland, and Jeremy~L. England.
\newblock Statistical {{Physics}} of {{Adaptation}}.
\newblock {\em Physical Review X}, 6(2):021036, June 2016.

\bibitem{jefferyStatisticalMechanicsLife2019}
Kate Jeffery, Robert Pollack, and Carlo Rovelli.
\newblock On the statistical mechanics of life: {{Schr}}\textbackslash "odinger
  revisited.
\newblock {\em arXiv:1908.08374 [physics]}, August 2019.

\bibitem{englandStatisticalPhysicsSelfreplication2013}
Jeremy~L. England.
\newblock Statistical physics of self-replication.
\newblock {\em The Journal of Chemical Physics}, 139(12):121923, August 2013.

\bibitem{skinnerImprovedBoundsEntropy2021}
Dominic~J. Skinner and J{\"o}rn Dunkel.
\newblock Improved bounds on entropy production in living systems.
\newblock {\em Proceedings of the National Academy of Sciences}, 118(18), May
  2021.

\bibitem{dunnLearningInferenceNonequilibrium2013}
Benjamin Dunn and Yasser Roudi.
\newblock Learning and inference in a nonequilibrium {{Ising}} model with
  hidden nodes.
\newblock {\em Physical Review E}, 87(2):022127, February 2013.

\bibitem{stillThermodynamicCostBenefit2020}
Susanne Still.
\newblock Thermodynamic {{Cost}} and {{Benefit}} of {{Memory}}.
\newblock {\em Physical Review Letters}, 124(5):050601, February 2020.

\bibitem{stillThermodynamicsPrediction2012}
Susanne Still, David~A. Sivak, Anthony~J. Bell, and Gavin~E. Crooks.
\newblock Thermodynamics of {{Prediction}}.
\newblock {\em Physical Review Letters}, 109(12):120604, September 2012.

\bibitem{ueltzhofferThermodynamicsPredictionDissipative2020}
Kai Ueltzh{\"o}ffer.
\newblock On the thermodynamics of prediction under dissipative adaptation.
\newblock {\em arXiv:2009.04006 [cond-mat, q-bio]}, September 2020.

\bibitem{kardesThermodynamicUncertaintyRelations2021}
G{\"u}lce Karde{\c s} and David~H. Wolpert.
\newblock Thermodynamic {{Uncertainty Relations}} for {{Multipartite
  Processes}}.
\newblock {\em arXiv:2101.01610 [cond-mat]}, March 2021.

\bibitem{wolpertMinimalEntropyProduction2020}
David~H. Wolpert.
\newblock Minimal entropy production rate of interacting systems.
\newblock {\em New Journal of Physics}, 22(11):113013, November 2020.

\bibitem{wolpertUncertaintyRelationsFluctuation2020}
David~H. Wolpert.
\newblock Uncertainty {{Relations}} and {{Fluctuation Theorems}} for {{Bayes
  Nets}}.
\newblock {\em Physical Review Letters}, 125(20):200602, November 2020.

\bibitem{crooksMarginalConditionalSecond2019}
Gavin~E. Crooks and Susanne Still.
\newblock Marginal and conditional second laws of thermodynamics.
\newblock {\em EPL (Europhysics Letters)}, 125(4):40005, March 2019.

\bibitem{horowitzThermodynamicsContinuousInformation2014}
Jordan~M. Horowitz and Massimiliano Esposito.
\newblock Thermodynamics with {{Continuous Information Flow}}.
\newblock {\em Physical Review X}, 4(3):031015, July 2014.

\bibitem{pougetInferenceComputationPopulation2003}
Alexandre Pouget, Peter Dayan, and Richard~S. Zemel.
\newblock Inference and computation with population codes.
\newblock {\em Annual Review of Neuroscience}, 26(1):381--410, March 2003.

\bibitem{knillBayesianBrainRole2004}
David~C. Knill and Alexandre Pouget.
\newblock The {{Bayesian}} brain: The role of uncertainty in neural coding and
  computation.
\newblock {\em Trends in Neurosciences}, 27(12):712--719, December 2004.

\bibitem{fristonFreeenergyPrincipleUnified2010}
Karl Friston.
\newblock The free-energy principle: A unified brain theory?
\newblock {\em Nature Reviews Neuroscience}, 11(2):127--138, February 2010.

\bibitem{raoPredictiveCodingVisual1999}
Rajesh P.~N. Rao and Dana~H. Ballard.
\newblock Predictive coding in the visual cortex: A functional interpretation
  of some extra-classical receptive-field effects.
\newblock {\em Nature Neuroscience}, 2(1):79--87, January 1999.

\bibitem{fristonActionBehaviorFreeenergy2010}
Karl~J. Friston, Jean Daunizeau, James Kilner, and Stefan~J. Kiebel.
\newblock Action and behavior: A free-energy formulation.
\newblock {\em Biological Cybernetics}, 102(3):227--260, March 2010.

\bibitem{fristonParcelsParticlesMarkov2020}
Karl~J. Friston, Erik~D. Fagerholm, Tahereh~S. Zarghami, Thomas Parr, In{\^e}s
  Hip{\'o}lito, Lo{\"i}c Magrou, and Adeel Razi.
\newblock Parcels and particles: {{Markov}} blankets in the brain.
\newblock {\em arXiv:2007.09704 [q-bio]}, July 2020.

\bibitem{fristonLifeWeKnow2013}
Karl Friston.
\newblock Life as we know it.
\newblock {\em Journal of The Royal Society Interface}, 10(86):20130475,
  September 2013.

\bibitem{parrMarkovBlanketsInformation2020}
Thomas Parr, Lancelot Da~Costa, and Karl Friston.
\newblock Markov blankets, information geometry and stochastic thermodynamics.
\newblock {\em Philosophical Transactions of the Royal Society A: Mathematical,
  Physical and Engineering Sciences}, 378(2164):20190159, February 2020.

\bibitem{fristonFreeEnergyPrinciple2019a}
Karl Friston.
\newblock A free energy principle for a particular physics.
\newblock {\em arXiv:1906.10184 [q-bio]}, June 2019.

\bibitem{fristonStochasticChaosMarkov2021}
Karl Friston, Conor Heins, Kai Ueltzh{\"o}ffer, Lancelot Da~Costa, and Thomas
  Parr.
\newblock Stochastic {{Chaos}} and {{Markov Blankets}}.
\newblock {\em Entropy}, 23(9):1220, September 2021.

\bibitem{wainwrightGraphicalModelsExponential2007}
Martin~J. Wainwright and Michael~I. Jordan.
\newblock Graphical {{Models}}, {{Exponential Families}}, and {{Variational
  Inference}}.
\newblock {\em Foundations and Trends\textregistered{} in Machine Learning},
  1(1\textendash 2):1--305, 2007.

\bibitem{eatonMultivariateStatisticsVector2007}
Morris~L. Eaton.
\newblock {\em Multivariate {{Statistics}}: {{A Vector Space Approach}}}.
\newblock {Institute of Mathematical Statistics}, 2007.

\bibitem{parrComputationalNeurologyActive2019}
Thomas Parr.
\newblock {\em The Computational Neurology of Active Vision}.
\newblock Ph.{{D}}. {{Thesis}}, University College London, {London}, 2019.

\bibitem{meisterNeuralCodeRetina1999}
Markus Meister and Michael~J. Berry.
\newblock The {{Neural Code}} of the {{Retina}}.
\newblock {\em Neuron}, 22(3):435--450, March 1999.

\bibitem{jamesGeneralisedInverse1978}
M.~James.
\newblock The generalised inverse.
\newblock {\em The Mathematical Gazette}, 62(420):109--114, June 1978.

\bibitem{aguileraHowParticularPhysics2021}
Miguel Aguilera, Beren Millidge, Alexander Tschantz, and Christopher~L.
  Buckley.
\newblock How particular is the physics of the {{Free Energy Principle}}?
\newblock {\em arXiv:2105.11203 [q-bio]}, May 2021.

\bibitem{mattinglyConvergenceNumericalTimeAveraging2010}
Jonathan~C. Mattingly, Andrew~M. Stuart, and M.~V. Tretyakov.
\newblock Convergence of {{Numerical Time}}-{{Averaging}} and {{Stationary
  Measures}} via {{Poisson Equations}}.
\newblock {\em SIAM Journal on Numerical Analysis}, 48(2):552--577, January
  2010.

\bibitem{rogersDiffusionsMarkovProcesses2000a}
L.~C.~G. Rogers and David Williams.
\newblock {\em Diffusions, {{Markov Processes}}, and {{Martingales}}:
  {{Volume}} 1: {{Foundations}}}, volume~1 of {\em Cambridge {{Mathematical
  Library}}}.
\newblock {Cambridge University Press}, {Cambridge}, second edition, 2000.

\bibitem{pavliotisStochasticProcessesApplications2014}
Grigorios~A. Pavliotis.
\newblock {\em Stochastic Processes and Applications: Diffusion Processes, the
  {{Fokker}}-{{Planck}} and {{Langevin}} Equations}.
\newblock Number volume 60 in Texts in Applied Mathematics. {Springer}, {New
  York}, 2014.

\bibitem{bierkensZigZagProcessSuperefficient2019}
Joris Bierkens, Paul Fearnhead, and Gareth Roberts.
\newblock The {{Zig}}-{{Zag}} process and super-efficient sampling for
  {{Bayesian}} analysis of big data.
\newblock {\em The Annals of Statistics}, 47(3):1288--1320, June 2019.

\bibitem{bierkensPiecewiseDeterministicScaling2017}
Joris Bierkens and Gareth Roberts.
\newblock A piecewise deterministic scaling limit of lifted
  {{Metropolis}}\textendash{{Hastings}} in the {{Curie}}\textendash{{Weiss}}
  model.
\newblock {\em The Annals of Applied Probability}, 27(2):846--882, April 2017.

\bibitem{bouchard-coteBouncyParticleSampler2018}
Alexandre {Bouchard-C{\^o}t{\'e}}, Sebastian~J. Vollmer, and Arnaud Doucet.
\newblock The {{Bouncy Particle Sampler}}: {{A Nonreversible Rejection}}-{{Free
  Markov Chain Monte Carlo Method}}.
\newblock {\em Journal of the American Statistical Association},
  113(522):855--867, April 2018.

\bibitem{rasmussenGaussianProcessesMachine2004}
Carl~Edward Rasmussen.
\newblock Gaussian {{Processes}} in {{Machine Learning}}.
\newblock In Olivier Bousquet, Ulrike {von Luxburg}, and Gunnar R{\"a}tsch,
  editors, {\em Advanced {{Lectures}} on {{Machine Learning}}: {{ML Summer
  Schools}} 2003, {{Canberra}}, {{Australia}}, {{February}} 2 - 14, 2003,
  {{T\"ubingen}}, {{Germany}}, {{August}} 4 - 16, 2003, {{Revised Lectures}}},
  Lecture {{Notes}} in {{Computer Science}}, pages 63--71. {Springer}, {Berlin,
  Heidelberg}, 2004.

\bibitem{arnoldRandomDynamicalSystems1998}
Ludwig Arnold.
\newblock {\em Random {{Dynamical Systems}}}.
\newblock Springer {{Monographs}} in {{Mathematics}}. {Springer-Verlag},
  {Berlin Heidelberg}, 1998.

\bibitem{biehlTechnicalCritiqueParts2021}
Martin Biehl, Felix~A. Pollock, and Ryota Kanai.
\newblock A {{Technical Critique}} of {{Some Parts}} of the {{Free Energy
  Principle}}.
\newblock {\em Entropy}, 23(3):293, March 2021.

\bibitem{fristonInterestingObservationsFree2021}
Karl~J. Friston, Lancelot Da~Costa, and Thomas Parr.
\newblock Some {{Interesting Observations}} on the {{Free Energy Principle}}.
\newblock {\em Entropy}, 23(8):1076, August 2021.

\bibitem{jafriGeneralizedSynchronyCoupled2016}
Haider~Hasan Jafri, R.~K.~Brojen Singh, and Ramakrishna Ramaswamy.
\newblock Generalized synchrony of coupled stochastic processes with
  multiplicative noise.
\newblock {\em Physical Review E}, 94(5):052216, November 2016.

\bibitem{cuminGeneralisingKuramotoModel2007}
D.~Cumin and C.~P. Unsworth.
\newblock Generalising the {{Kuramoto}} model for the study of neuronal
  synchronisation in the brain.
\newblock {\em Physica D: Nonlinear Phenomena}, 226(2):181--196, February 2007.

\bibitem{palaciosEmergenceSynchronyNetworks2019}
Ensor~Rafael Palacios, Takuya Isomura, Thomas Parr, and Karl Friston.
\newblock The emergence of synchrony in networks of mutually inferring neurons.
\newblock {\em Scientific Reports}, 9(1):6412, April 2019.

\bibitem{bartlettReconstructionGaussianQuantum2012}
Stephen~D. Bartlett, Terry Rudolph, and Robert~W. Spekkens.
\newblock Reconstruction of {{Gaussian}} quantum mechanics from {{Liouville}}
  mechanics with an epistemic restriction.
\newblock {\em Physical Review A}, 86(1):012103, July 2012.

\bibitem{kullbackInformationSufficiency1951}
S.~Kullback and R.~A. Leibler.
\newblock On {{Information}} and {{Sufficiency}}.
\newblock {\em The Annals of Mathematical Statistics}, 22(1):79--86, March
  1951.

\bibitem{bogaczTutorialFreeenergyFramework2017}
Rafal Bogacz.
\newblock A tutorial on the free-energy framework for modelling perception and
  learning.
\newblock {\em Journal of Mathematical Psychology}, 76:198--211, February 2017.

\bibitem{fristonPredictiveCodingFreeenergy2009}
Karl Friston and Stefan Kiebel.
\newblock Predictive coding under the free-energy principle.
\newblock {\em Philosophical Transactions of the Royal Society B: Biological
  Sciences}, 364(1521):1211--1221, May 2009.

\bibitem{chaoLargeScaleCorticalNetworks2018}
Zenas~C. Chao, Kana Takaura, Liping Wang, Naotaka Fujii, and Stanislas Dehaene.
\newblock Large-{{Scale Cortical Networks}} for {{Hierarchical Prediction}} and
  {{Prediction Error}} in the {{Primate Brain}}.
\newblock {\em Neuron}, 100(5):1252--1266.e3, May 2018.

\bibitem{iglesiasHierarchicalPredictionErrors2013}
Sandra Iglesias, Christoph Mathys, Kay~H. Brodersen, Lars Kasper, Marco
  Piccirelli, Hanneke E.~M. {den Ouden}, and Klaas~E. Stephan.
\newblock Hierarchical {{Prediction Errors}} in {{Midbrain}} and {{Basal
  Forebrain}} during {{Sensory Learning}}.
\newblock {\em Neuron}, 80(2):519--530, October 2013.

\bibitem{dawModelBasedInfluencesHumans2011}
Nathaniel~D. Daw, Samuel~J. Gershman, Ben Seymour, Peter Dayan, and Raymond~J.
  Dolan.
\newblock Model-{{Based Influences}} on {{Humans}}' {{Choices}} and {{Striatal
  Prediction Errors}}.
\newblock {\em Neuron}, 69(6):1204--1215, March 2011.

\bibitem{ottobreMarkovChainMonte2016}
Michela Ottobre.
\newblock Markov {{Chain Monte Carlo}} and {{Irreversibility}}.
\newblock {\em Reports on Mathematical Physics}, 77:267--292, June 2016.

\bibitem{rey-belletIrreversibleLangevinSamplers2015}
Luc {Rey-Bellet} and Kostantinos Spiliopoulos.
\newblock Irreversible {{Langevin}} samplers and variance reduction: A large
  deviation approach.
\newblock {\em Nonlinearity}, 28(7):2081--2103, July 2015.

\bibitem{bleiVariationalInferenceReview2017}
David~M. Blei, Alp Kucukelbir, and Jon~D. McAuliffe.
\newblock Variational {{Inference}}: {{A Review}} for {{Statisticians}}.
\newblock {\em Journal of the American Statistical Association},
  112(518):859--877, April 2017.

\bibitem{buckleyFreeEnergyPrinciple2017}
Christopher~L. Buckley, Chang~Sub Kim, Simon McGregor, and Anil~K. Seth.
\newblock The free energy principle for action and perception: {{A}}
  mathematical review.
\newblock {\em Journal of Mathematical Psychology}, 81:55--79, December 2017.

\bibitem{bealVariationalAlgorithmsApproximate2003}
Matthew~James Beal.
\newblock {\em Variational {{Algorithms}} for {{Approximate Bayesian
  Inference}}}.
\newblock Ph.{{D}}. {{Thesis}}, University of London, 2003.

\bibitem{schwartzSpiketriggeredNeuralCharacterization2006}
Odelia Schwartz, Jonathan~W. Pillow, Nicole~C. Rust, and Eero~P. Simoncelli.
\newblock Spike-triggered neural characterization.
\newblock {\em Journal of Vision}, 6(4):484--507, July 2006.

\bibitem{sayerTimeCourseAmplitude1990}
R.~J. Sayer, M.~J. Friedlander, and S.~J. Redman.
\newblock The time course and amplitude of {{EPSPs}} evoked at synapses between
  pairs of {{CA3}}/{{CA1}} neurons in the hippocampal slice.
\newblock {\em The Journal of Neuroscience: The Official Journal of the Society
  for Neuroscience}, 10(3):826--836, March 1990.

\bibitem{luckIntroductionEventRelatedPotential2014}
Steven~J. Luck.
\newblock {\em An {{Introduction}} to the {{Event}}-{{Related Potential
  Technique}}}.
\newblock {A Bradford Book}, {Cambridge, MA, USA}, second edition, May 2014.

\bibitem{ueltzhofferDeepActiveInference2018}
Kai Ueltzh{\"o}ffer.
\newblock Deep {{Active Inference}}.
\newblock {\em Biological Cybernetics}, 112(6):547--573, December 2018.

\bibitem{millidgeDeepActiveInference2020}
Beren Millidge.
\newblock Deep active inference as variational policy gradients.
\newblock {\em Journal of Mathematical Psychology}, 96:102348, June 2020.

\bibitem{heinsDeepActiveInference2020}
R.~Conor Heins, M.~Berk Mirza, Thomas Parr, Karl Friston, Igor Kagan, and
  Arezoo Pooresmaeili.
\newblock Deep {{Active Inference}} and {{Scene Construction}}.
\newblock {\em Frontiers in Artificial Intelligence}, 3:81, 2020.

\bibitem{lanillosRobotSelfOther2020}
Pablo Lanillos, Jordi Pages, and Gordon Cheng.
\newblock Robot self/other distinction: Active inference meets neural networks
  learning in a mirror.
\newblock In {\em European {{Conference}} on {{Artificial Intelligence}}}. {IOS
  press}, April 2020.

\bibitem{verbelenActiveInferenceFirst2020}
Tim Verbelen, Pablo Lanillos, Christopher Buckley, and Cedric~De Boom, editors.
\newblock {\em Active {{Inference}}: {{First International Workshop}}, {{IWAI}}
  2020, {{Co}}-Located with {{ECML}}/{{PKDD}} 2020, {{Ghent}}, {{Belgium}},
  {{September}} 14, 2020, {{Proceedings}}}.
\newblock Communications in {{Computer}} and {{Information Science}}. {Springer
  International Publishing}, 2020.

\bibitem{adamsPredictionsNotCommands2013}
Rick~A. Adams, Stewart Shipp, and Karl~J. Friston.
\newblock Predictions not commands: Active inference in the motor system.
\newblock {\em Brain Structure \& Function}, 218(3):611--643, May 2013.

\bibitem{pezzatoNovelAdaptiveController2020}
Corrado Pezzato, Riccardo Ferrari, and Carlos~Hern{\'a}ndez Corbato.
\newblock A {{Novel Adaptive Controller}} for {{Robot Manipulators Based}} on
  {{Active Inference}}.
\newblock {\em IEEE Robotics and Automation Letters}, 5(2):2973--2980, April
  2020.

\bibitem{oliverEmpiricalStudyActive2021}
Guillermo Oliver, Pablo Lanillos, and Gordon Cheng.
\newblock An empirical study of active inference on a humanoid robot.
\newblock {\em IEEE Transactions on Cognitive and Developmental Systems}, pages
  1--1, 2021.

\bibitem{lelievreOptimalNonreversibleLinear2013}
Tony Leli{\`e}vre, Francis Nier, and Grigorios~A. Pavliotis.
\newblock Optimal non-reversible linear drift for the convergence to
  equilibrium of a diffusion.
\newblock {\em Journal of Statistical Physics}, 152(2):237--274, July 2013.

\bibitem{koudahlWorkedExampleFokkerPlanckBased2020}
Magnus~T. Koudahl and Bert {de Vries}.
\newblock A {{Worked Example}} of {{Fokker}}-{{Planck}}-{{Based Active
  Inference}}.
\newblock In Tim Verbelen, Pablo Lanillos, Christopher~L. Buckley, and Cedric
  De~Boom, editors, {\em Active {{Inference}}}, Communications in {{Computer}}
  and {{Information Science}}, pages 28--34, {Cham}, 2020. {Springer
  International Publishing}.

\bibitem{fristonWhatOptimalMotor2011}
Karl Friston.
\newblock What {{Is Optimal}} about {{Motor Control}}?
\newblock {\em Neuron}, 72(3):488--498, November 2011.

\bibitem{sancaktarEndtoEndPixelBasedDeep2020}
Cansu Sancaktar, Marcel {van Gerven}, and Pablo Lanillos.
\newblock End-to-{{End Pixel}}-{{Based Deep Active Inference}} for {{Body
  Perception}} and {{Action}}.
\newblock {\em arXiv:2001.05847 [cs, q-bio]}, May 2020.

\bibitem{baltieriPIDControlProcess2019}
Manuel Baltieri and Christopher~L. Buckley.
\newblock {{PID Control}} as a {{Process}} of {{Active Inference}} with
  {{Linear Generative Models}}.
\newblock {\em Entropy}, 21(3):257, March 2019.

\bibitem{kosinski2008literature}
Robert~J Kosinski.
\newblock A literature review on reaction time.
\newblock {\em Clemson University}, 10(1), 2008.

\bibitem{roskillyMarineSystemsIdentification2015}
Tony Roskilly and Dr~Rikard Mikalsen.
\newblock {\em Marine {{Systems Identification}}, {{Modeling}} and
  {{Control}}}.
\newblock {Butterworth-Heinemann}, {Amsterdam ; Boston}, illustrated edition
  edition, March 2015.

\bibitem{astromPidControllers1995}
Karl~Johan {\AA}str{\"o}m.
\newblock {\em Pid {{Controllers}}}.
\newblock {International Society for Measurement and Control}, January 1995.

\bibitem{mitterTheoryNonlinearStochastic1981}
Sanjoy Mitter, Giorgio Picci, and Anders Lindquist.
\newblock Toward a theory of nonlinear stochastic realization.
\newblock In {\em Feedback and {{Synthesis}} of {{Linear}} and {{Nonlinear
  Systems}}}, 1981.

\bibitem{lindquistLinearStochasticSystems2015}
Anders Lindquist and Giorgio Picci.
\newblock {\em Linear {{Stochastic Systems}}: {{A Geometric Approach}} to
  {{Modeling}}, {{Estimation}} and {{Identification}}}.
\newblock Series in {{Contemporary Mathematics}}. {Springer-Verlag}, {Berlin
  Heidelberg}, 2015.

\bibitem{lindquistRealizationTheoryMultivariate1985}
Anders Lindquist and Giorgio Picci.
\newblock Realization {{Theory}} for {{Multivariate Stationary Gaussian
  Processes}}.
\newblock {\em SIAM Journal on Control and Optimization}, 23(6):809--857,
  November 1985.

\bibitem{doobBrownianMovementStochastic1942}
J.~L. Doob.
\newblock The {{Brownian Movement}} and {{Stochastic Equations}}.
\newblock {\em Annals of Mathematics}, 43(2):351--369, 1942.

\bibitem{wangTheoryBrownianMotion2014}
Ming~Chen Wang and G.~E. Uhlenbeck.
\newblock On the {{Theory}} of the {{Brownian Motion II}}.
\newblock In {\em Selected {{Papers}} on {{Noise}} and {{Stochastic
  Processes}}}. {Dover}, 2014.

\bibitem{rey-belletOpenClassicalSystems2006}
Luc {Rey-Bellet}.
\newblock Open {{Classical Systems}}.
\newblock In St{\'e}phane Attal, Alain Joye, and Claude-Alain Pillet, editors,
  {\em Open {{Quantum Systems II}}: {{The Markovian Approach}}}, Lecture
  {{Notes}} in {{Mathematics}}, pages 41--78. {Springer}, {Berlin, Heidelberg},
  2006.

\bibitem{kryachkovFinitetimeStabilizationIntegrator2010}
Mikhail Kryachkov, Andrey Polyakov, and Vadim Strygin.
\newblock Finite-time stabilization of an integrator chain using only signs of
  the state variables.
\newblock In {\em 2010 11th {{International Workshop}} on {{Variable Structure
  Systems}} ({{VSS}})}, pages 510--515, June 2010.

\bibitem{zimenkoFinitetimeFixedtimeStabilization2018}
Konstantin Zimenko, Andrey Polyakov, Denis Efimo, and Wilfrid Perruquetti.
\newblock Finite-time and fixed-time stabilization for integrator chain of
  arbitrary order*.
\newblock In {\em 2018 {{European Control Conference}} ({{ECC}})}, pages
  1631--1635, June 2018.

\bibitem{fristonVariationalFiltering2008}
K.~J. Friston.
\newblock Variational filtering.
\newblock {\em NeuroImage}, 41(3):747--766, July 2008.

\bibitem{fristonVariationalTreatmentDynamic2008}
K.~J. Friston, N.~{Trujillo-Barreto}, and J.~Daunizeau.
\newblock {{DEM}}: A variational treatment of dynamic systems.
\newblock {\em NeuroImage}, 41(3):849--885, July 2008.

\bibitem{fristonGeneralisedFiltering2010}
Karl Friston, Klaas Stephan, Baojuan Li, and Jean Daunizeau.
\newblock Generalised {{Filtering}}.
\newblock {\em Mathematical Problems in Engineering}, 2010:1--34, 2010.

\bibitem{parrComputationalNeurologyMovement2021}
Thomas Parr, Jakub Limanowski, Vishal Rawji, and Karl Friston.
\newblock The computational neurology of movement under active inference.
\newblock {\em Brain}, (awab085), March 2021.

\bibitem{gomesMeanFieldLimits2020}
S.~N. Gomes, G.~A. Pavliotis, and U.~Vaes.
\newblock Mean {{Field Limits}} for {{Interacting Diffusions}} with {{Colored
  Noise}}: {{Phase Transitions}} and {{Spectral Numerical Methods}}.
\newblock {\em Multiscale Modeling \& Simulation}, 18(3):1343--1370, January
  2020.

\bibitem{tayorNonlinearStochasticRealization1989}
T.~J.~S. Tayor and M.~Pavon.
\newblock On the nonlinear stochastic realization problem.
\newblock {\em Stochastics and Stochastic Reports}, 26(2):65--79, February
  1989.

\bibitem{frazhoStochasticRealizationTheory1982}
A.~E. Frazho.
\newblock On stochastic realization theory.
\newblock {\em Stochastics}, 7(1-2):1--27, January 1982.

\bibitem{fristonGraphicalBrainBelief2017}
Karl~J. Friston, Thomas Parr, and Bert {de Vries}.
\newblock The graphical brain: {{Belief}} propagation and active inference.
\newblock {\em Network Neuroscience}, 1(4):381--414, December 2017.

\bibitem{chevalierDesignAnalysisProportionalIntegralDerivative2019}
Michael Chevalier, Mariana {G{\'o}mez-Schiavon}, Andrew~H. Ng, and Hana
  {El-Samad}.
\newblock Design and {{Analysis}} of a
  {{Proportional}}-{{Integral}}-{{Derivative Controller}} with {{Biological
  Molecules}}.
\newblock {\em Cell Systems}, 9(4):338--353.e10, October 2019.

\bibitem{yiRobustPerfectAdaptation2000}
Tau-Mu Yi, Yun Huang, Melvin~I. Simon, and John Doyle.
\newblock Robust perfect adaptation in bacterial chemotaxis through integral
  feedback control.
\newblock {\em Proceedings of the National Academy of Sciences},
  97(9):4649--4653, April 2000.

\bibitem{fristonHierarchicalModelsBrain2008}
Karl Friston.
\newblock Hierarchical {{Models}} in the {{Brain}}.
\newblock {\em PLoS Computational Biology}, 4(11):e1000211, November 2008.

\bibitem{kalmanNewApproachLinear1960}
R.~E. Kalman.
\newblock A {{New Approach}} to {{Linear Filtering}} and {{Prediction
  Problems}}.
\newblock {\em Journal of Basic Engineering}, 82(1):35--45, March 1960.

\bibitem{fristonVariationalFreeEnergy2007}
Karl Friston, J{\'e}r{\'e}mie Mattout, Nelson {Trujillo-Barreto}, John
  Ashburner, and Will Penny.
\newblock Variational free energy and the {{Laplace}} approximation.
\newblock {\em NeuroImage}, 34(1):220--234, January 2007.

\bibitem{fristonTheoryCorticalResponses2005}
Karl Friston.
\newblock A theory of cortical responses.
\newblock {\em Philosophical Transactions of the Royal Society B: Biological
  Sciences}, 360(1456):815--836, April 2005.

\bibitem{pezzuloActiveInferenceView2012}
Giovanni Pezzulo.
\newblock An {{Active Inference}} view of cognitive control.
\newblock {\em Frontiers in Psychology}, 3, 2012.

\bibitem{pelletUsingMarkovBlankets2008}
Jean-Philippe Pellet and Andr{\'e} Elisseeff.
\newblock Using {{Markov Blankets}} for {{Causal Structure Learning}}.
\newblock {\em Journal of Machine Learning Research}, 9(43):1295--1342, 2008.

\bibitem{tzenNeuralStochasticDifferential2019}
Belinda Tzen and M.~Raginsky.
\newblock Neural {{Stochastic Differential Equations}}: {{Deep Latent Gaussian
  Models}} in the {{Diffusion Limit}}.
\newblock {\em ArXiv}, 2019.

\bibitem{hornMatrixAnalysisSecond2012}
Roger~A. Horn.
\newblock {\em Matrix {{Analysis}}: {{Second Edition}}}.
\newblock {Cambridge University Press}, {New York, NY}, 2nd edition edition,
  December 2012.

\bibitem{rogersDiffusionsMarkovProcesses2000}
L.~C.~G. Rogers and David Williams.
\newblock {\em Diffusions, {{Markov Processes}} and {{Martingales}}: {{Volume}}
  2: {{It\^o Calculus}}}, volume~2 of {\em Cambridge {{Mathematical Library}}}.
\newblock {Cambridge University Press}, {Cambridge}, second edition, 2000.

\bibitem{oksendalStochasticDifferentialEquations2003}
Bernt {\O}ksendal.
\newblock {\em Stochastic {{Differential Equations}}: {{An Introduction}} with
  {{Applications}}}.
\newblock Universitext. {Springer-Verlag}, {Berlin Heidelberg}, sixth edition,
  2003.

\bibitem{grahamCovariantFormulationNonequilibrium1977}
Robert Graham.
\newblock Covariant formulation of non-equilibrium statistical thermodynamics.
\newblock {\em Zeitschrift f\"ur Physik B Condensed Matter}, 26(4):397--405,
  December 1977.

\bibitem{eyinkHydrodynamicsFluctuationsOutside1996}
Gregory~L. Eyink, Joel~L. Lebowitz, and Herbert Spohn.
\newblock Hydrodynamics and fluctuations outside of local equilibrium:
  {{Driven}} diffusive systems.
\newblock {\em Journal of Statistical Physics}, 83(3):385--472, May 1996.

\bibitem{aoPotentialStochasticDifferential2004}
P.~Ao.
\newblock Potential in stochastic differential equations: Novel construction.
\newblock {\em Journal of Physics A: Mathematical and General}, 37(3):L25--L30,
  January 2004.

\bibitem{qianDecompositionIrreversibleDiffusion2013}
Hong Qian.
\newblock A decomposition of irreversible diffusion processes without detailed
  balance.
\newblock {\em Journal of Mathematical Physics}, 54(5):053302, May 2013.

\bibitem{maCompleteRecipeStochastic2015}
Yi-An Ma, Tianqi Chen, and Emily~B. Fox.
\newblock A {{Complete Recipe}} for {{Stochastic Gradient MCMC}}.
\newblock {\em arXiv:1506.04696 [math, stat]}, October 2015.

\bibitem{barpUnifyingCanonicalDescription2021}
Alessandro Barp, So~Takao, Michael Betancourt, Alexis Arnaudon, and Mark
  Girolami.
\newblock A {{Unifying}} and {{Canonical Description}} of
  {{Measure}}-{{Preserving Diffusions}}.
\newblock {\em arXiv:2105.02845 [math, stat]}, May 2021.

\bibitem{chaudhariStochasticGradientDescent2018}
Pratik Chaudhari and Stefano Soatto.
\newblock Stochastic gradient descent performs variational inference, converges
  to limit cycles for deep networks.
\newblock In {\em International {{Conference}} on {{Learning
  Representations}}}, February 2018.

\bibitem{daiLargeBatchTraining2020}
Xiaowu Dai and Yuhua Zhu.
\newblock On {{Large Batch Training}} and {{Sharp Minima}}: {{A
  Fokker}}\textendash{{Planck Perspective}}.
\newblock {\em Journal of Statistical Theory and Practice}, 14(3):53, July
  2020.

\bibitem{aokiEntropyProductionLiving1995}
Ichiro Aoki.
\newblock Entropy production in living systems: From organisms to ecosystems.
\newblock {\em Thermochimica Acta}, 250(2):359--370, February 1995.

\bibitem{haussmannTimeReversalDiffusions1986}
U.~G. Haussmann and E.~Pardoux.
\newblock Time {{Reversal}} of {{Diffusions}}.
\newblock {\em Annals of Probability}, 14(4):1188--1205, October 1986.

\bibitem{riskenFokkerPlanckEquationMethods1996}
Hannes Risken and Till Frank.
\newblock {\em The {{Fokker}}-{{Planck Equation}}: {{Methods}} of {{Solution}}
  and {{Applications}}}.
\newblock Springer {{Series}} in {{Synergetics}}. {Springer-Verlag}, {Berlin
  Heidelberg}, second edition, 1996.

\bibitem{RealAnalysisEvery}
Real analysis - {{Every}} divergence-free vector field generated from
  skew-symmetric matrix.
\newblock
  https://math.stackexchange.com/questions/578898/every-divergence-free-vector-field-generated-from-skew-symmetric-matrix.

\end{thebibliography}

\end{document}